\documentclass[aps,prb,twocolumn]{revtex4-2}
\usepackage{graphicx}  
\usepackage{amsmath,amssymb}  
\usepackage{hyperref}
\hypersetup{
	colorlinks=true,
	linkcolor=blue,
	citecolor=blue,
	urlcolor=blue
}
\usepackage{float}  
\usepackage{dutchcal}  
\usepackage{titlesec}  
﻿

\begin{document}

	\title{Eigenstate Thermalization and Spectral Imprints of the Hamiltonian in Local Observables}
	\author{Shivam~Mishra}
	\email{shivam.2023rph01@mnnit.ac.in}
	
	\author{C~Jisha}
	\email{jisha.2021rph03@mnnit.ac.in}
	
	\author{Ravi~Prakash}
	\email{ravi.prakash@mnnit.ac.in}
	
	\affiliation{Department of Physics, Motilal Nehru National Institute of Technology (MNNIT), Prayagraj--211004, India}

﻿

\begin{abstract}
	The Eigenstate Thermalization Hypothesis explains thermalization in isolated quantum systems through the statistical properties of observables in the energy eigenbasis. We investigate the crossover from integrability to chaos in the spin-$1/2$ XXZ chain, establishing a direct correspondence between the spectral correlations of the Hamiltonian and local observables expressed in the energy eigenbasis as a signature of ergodicity breaking. By introducing a local perturbation that drives the system from integrability to chaos, we track the standard ETH indicators and the eigenstate entanglement entropy. We introduce a submatrix-based framework for analyzing local observables in the energy eigenbasis. By extracting real-symmetric blocks along the diagonal of the local observables represented in eigenbasis, we show that these submatrices exhibit both the short-range and long-range spectral features of the Hamiltonian. Remarkably, this correspondence persists even in a partially ergodic regime, indicating that the emergence of chaos is already encoded locally within the observables' matrix structure and that small blocks are sufficient to capture the underlying spectral correlations.
	
\end{abstract}
﻿
\maketitle

\section{Introduction}
﻿Since the birth of quantum mechanics, a key question has endured: how do isolated quantum systems achieve thermal equilibrium under a unitary evolution? This question was raised in the early 20th century \cite{1,2,3,4}, and has taken on new importance with recent ultracold atom experiments \cite{5,6,7}. These experiments, with their excellent isolation from the environment, allow many-body systems to evolve coherently over long periods. They show that chaotic quantum systems thermalize \cite{8,9,10,11}, whereas integrable systems, constrained by an extensive set of conserved quantities, fail to thermalize \cite{12,13,14,15}. The Eigenstate Thermalization Hypothesis (ETH) provides a foundation to understand how typical isolated quantum systems achieve thermal equilibrium \cite{5,17,18,19,20}.
﻿
The ETH ansatz describes the matrix elements of a local operator $Z$ in the eigenbasis of the Hamiltonian $H$ \cite{5,16,17,18,20},
\begin{align}
	\langle \alpha | Z | \beta \rangle = \bar{Z}(\bar{E}) \delta_{\alpha \beta} 
	+ e^{-S(\bar{E})/2} f(\bar{E}, \omega) R_{\alpha \beta},
	\label{Eq:ETH}
\end{align}
where, $|\alpha\rangle$ and $|\beta\rangle$ represent the eigenvectors corresponding to eigenvalues $E_\alpha$ and $E_\beta$ respectively, $\bar{E} = (E_\alpha + E_\beta)/2$, $\omega = E_\alpha - E_\beta$, $S(\bar{E})$ represents the thermodynamic entropy, \( \bar{Z}(\bar{E}) \) and \( f(\bar{E}, \omega) \) are smooth functions of their arguments, and \( R_{\alpha \beta} \) are random variables with zero mean and unit variance \cite{21,22,23,24}. In chaotic systems, diagonal elements \( Z_{\alpha \alpha} \) vary smoothly with energy \cite{23,24,25} and their fluctuations decay exponentially with system size \cite{24,25}, while off-diagonal elements \( Z_{\alpha \beta} \) (\( \alpha \neq \beta \)) follow Gaussian distribution \cite{23,24,28,29,30}, supporting thermalization consistent with the fluctuation-dissipation theorem \cite{26,27}. However, integrable systems violate ETH and exhibit persistent fluctuations between eigenstates and non-Gaussian statistics of the off-diagonal elements \cite{29,30}.

The spin-\(\tfrac{1}{2}\) XXZ chain is a well-established model for exploring the crossover from integrability to chaos in its finite-size limit \cite{31,32,33}. 
In the integrable regime, its nearest-neighbor spacing distribution (NNSD) follows the Poisson distribution as $P(s) = e^{-s}$, indicating uncorrelated eigenvalues \cite{34,35,36,37,38,39}. 
Introducing a local perturbation can break integrability and drive the system toward the Wigner distribution as $P(s) = \frac{\pi s}{2} e^{-\pi s^2/4}$, characterized by level repulsion and spectral correlations typical of chaotic systems \cite{25,34,35,36,37,38,39,40,41}.

Although the ETH effectively describes thermalization in fully chaotic systems and its absence in integrable ones~\cite{24,25,29}, less attention has been given to how thermalization and ergodicity emerge gradually across the integrability to chaos crossover. Previous studies of the crossover from integrability to chaos have primarily focused on transitions in spectral correlations, which are studied through nearest-neighbour spacing distributions, spectral rigidity, or number-variance statistics of eigenenergies \cite{34,35,36,37,38,39,40}.

However, the statistical properties of matrix elements of local operators in the energy eigenbasis \cite{25,29,30,40}, and the behavior of eigenstate entanglement entropy \cite{30,42,43,44}, are discussed only in the integrable and fully chaotic limits, not in the crossover regime. These properties are central to understanding how quantum chaos governs equilibration; however, despite their importance, the evolution of these quantities across the integrability to chaos crossover remains unclear. It is an interesting and important problem to analyze the validity of ETH for weakly chaotic systems that are not fully ergodic, at the level of local operator matrix elements, in order to understand how correlations emerge within the eigenstate structure of many-body systems in the crossover regime.

\begin{figure}[htbp]
	\centering
	
	\includegraphics[width=\linewidth]{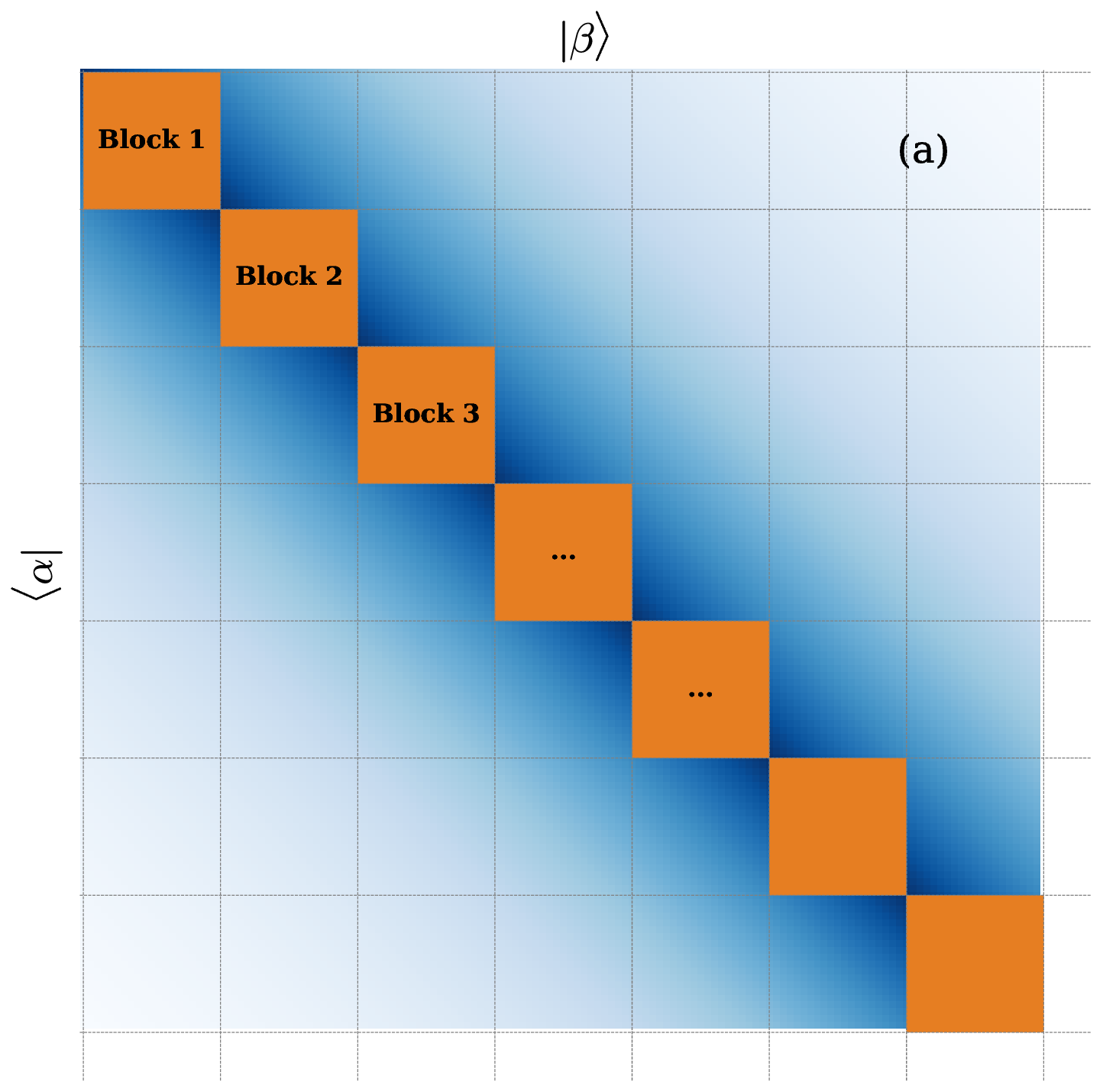}
	
	\includegraphics[width=\linewidth]{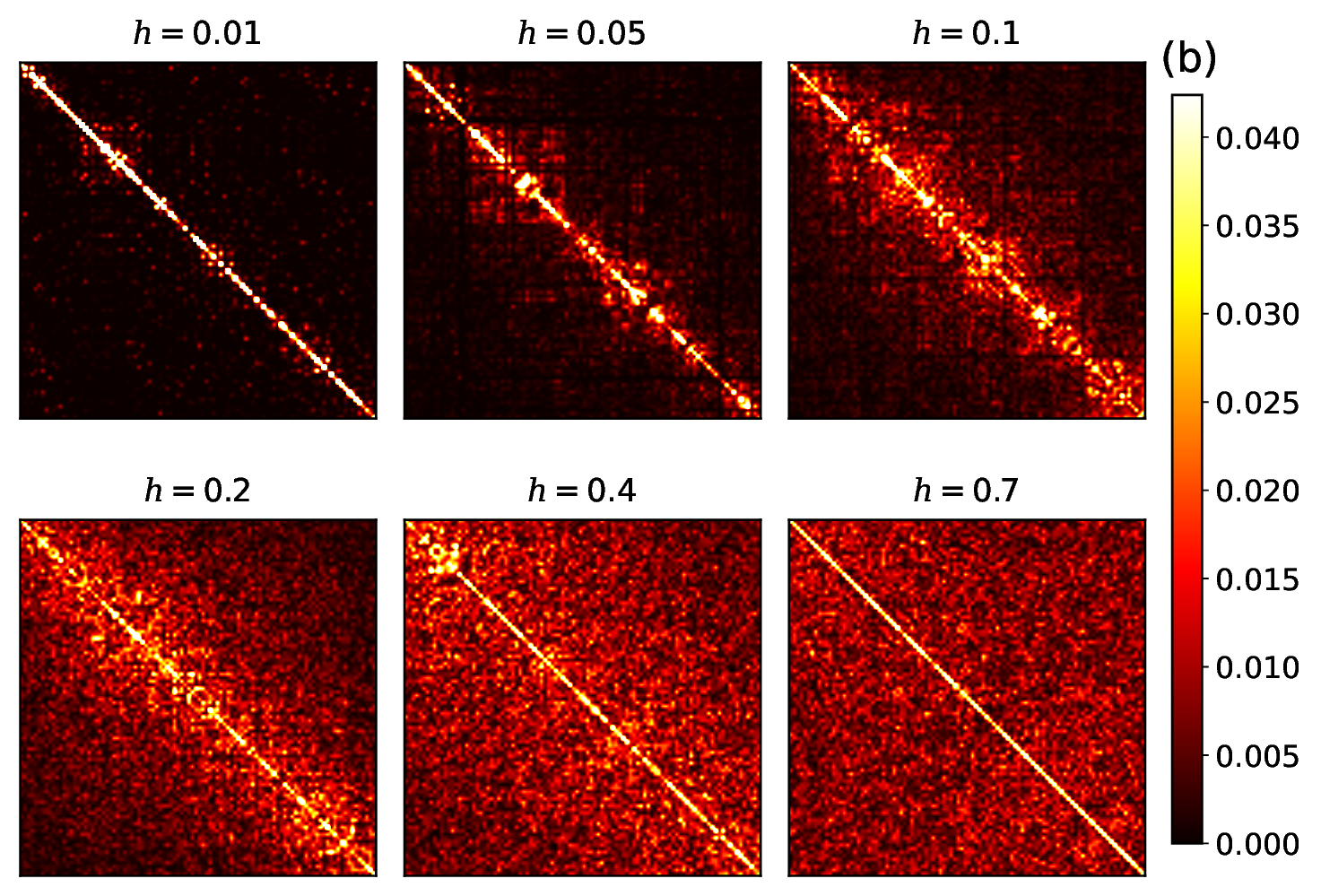}
	﻿
	\caption{(a) Schematic representation of submatrices of a local observable in the eigenbasis of the Hamiltonian, where each block corresponds to an energy-resolved window and captures correlations among matrix elements within that localized sector. (b) Density plot of a representative block of the operator $O$. Qualitatively similar structure is observed for blocks taken across the diagonal for both operators [see Eq. (\ref{eq:local_op1}) and (\ref{eq:local_op2}) ].}
	\label{fig:subm}
\end{figure}

To study this gap, we introduce a complementary framework rooted in the ETH, which directly probes the statistical structure of operator matrix elements represented in the eigenbasis. In this framework, we construct submatrices from the representation of an observable in the eigenbasis of the Hamiltonian [see Fig. \ref{fig:subm} (a)]. The individual blocks represent a similar qualitative structure as that of a power-law banded random matrix (PLBRM) [see Fig. \ref{fig:subm} (b)], see what PLBRM is in Refs.\cite{PLBRMa, rao2022plrbm}. We observed that the spectral correlations of the submatrices closely mirror those of the full Hamiltonian.
This correspondence holds not only in the fully chaotic regime but also across the crossover, where the system remains only partially ergodic. 

Furthermore, the spectra of these submatrices preserve both the short- and long-range spectral correlations, reflecting the universal ETH statistics even within reduced operator blocks. 
Overall, this submatrix-based framework offers a detailed insight into how matrix element correlations in the energy eigenbasis are progressively established within the local operators of isolated many-body systems.

In this article, we investigate the integrability to chaos crossover in the spin-$1/2$ XXZ chain, where non-integrability is introduced by adding a local perturbation on one of the spins. We analyze several diagnostics of chaos, \textit{e.g.}, spectral correlations (both short- and long-range), statistical properties of matrix elements of the observables (including the smoothness of diagonal elements, the Gaussianity of off-diagonal elements, and the exponential variance decay of off-diagonal elements with energy difference $\omega$), and eigenstate entanglement entropy. We introduce a submatrix framework showing that the spectral correlations of the  Hamiltonian are encoded in the operator structure itself. We show that this correspondence is not model-specific, and we further verify it for another many-body interacting system, demonstrating the generality of our result [see Appendix. \ref{app:bh_spacing_ratio}].
﻿
﻿

This paper is organized as follows. 
Section~\ref{sec:model} introduces an example of a many-body interacting system by the XXZ Hamiltonian. 
Section~\ref{sec:spectral_correlations}  presents numerical results for spectral correlations across the integrability to chaos crossover. 
Section~\ref{sec:eth} examines standard ETH diagnostics across the crossover regime.
Section~\ref{sec:submatrix} describes the submatrix framework and its correspondence with Hamiltonian spectral statistics. 
Finally, Section~\ref{sec:summary} provides a brief summary of our findings.

\section{XXZ Spin Chain}
\label{sec:model}
We consider the Heisenberg (XXZ) spin-$1/2$ chain with open boundary conditions, where nonintegrability is introduced by a local perturbation applied at site $(L/2 - 1)$, with $L$ representing the total number of sites. The Hamiltonian is given by,
﻿

\begin{equation}
	\begin{aligned}
		H &= H_0 + V, \\
		H_0 &= \sum_{i=1}^{L-1} J \left( S_i^x S_{i+1}^x + S_i^y S_{i+1}^y + \Delta S_i^z S_{i+1}^z \right), \\
		V &= h \left( S^z_{\frac{L}{2}-1} + S^x_{\frac{L}{2}-1} \right)
	\end{aligned}
	\label{eq:perturbation}
\end{equation}

where \( S_i^{x,y,z} \) are spin-$1/2$ operators (considering $\hbar =1$), $J = 1.0$ is the coupling strength, \( \Delta = \frac{\pi}{4} \) is the anisotropy parameter, and $h$ is the strength of the perturbation applied equally in the $x$- and $z$-directions. We consider a chain of $L=14$ spins and vary the perturbation strength as $h=\{0.01,\,0.05,\,0.1,\,0.2,\,0.4,\,0.7\}$ to induce a smooth crossover from integrability to chaos. As the system size increases, a progressively weaker perturbation is sufficient for the onset of chaos.

﻿
\section{Spectral Correlations}
\label{sec:spectral_correlations}

We begin with the evaluation of spectral correlations for the Hamiltonian Eq.~(\ref{eq:perturbation}). The short- and long-range correlations are evaluated using NNSD and level number variance, as discussed below.

\subsection{Nearest-Neighbor Spacing Distribution and the Number Variance}
The fluctuation statistics can be measured only after filtering the global smooth variation of spectral density. The unfolding procedure rescales the spectrum so that the local average density (or local average spacing) becomes uniform (preferably one). The unfolded spectrum thus has uniform average density. We characterize the crossover from integrability to chaos by examining both short- and long-range spectral correlations of the unfolded spectrum.

Short-range correlations are characterized through the NNSD. We compute the spacings between consecutive unfolded energy eigenvalues, $s_i = E_{i+1} - E_i$, where \(E_i\) are the ordered and unfolded eigenvalues. Intermediate spectral statistics across the integrability to chaos crossover are often described by the Brody distribution \cite{49,50,51,52,53}:
\begin{align}
	P(s) = (\gamma + 1) b s^\gamma \exp(-b s^{\gamma + 1}),
\end{align}
where $b = [ \Gamma ((\gamma + 2)/(\gamma + 1)) ]^{\gamma + 1}$ is the normalization constant with $\Gamma$ representing the Euler's gamma function. The parameter $0 \le \gamma \le1$ represents the crossover with $\gamma = 0$ and $1$ representing the Poisson and Wigner spacing distributions, respectively. 
﻿	

The level number variance statistic captures long-range correlations \cite{36,38,53}, which quantifies the rigidity of the spectrum by partitioning energy levels into multiple intervals of length \(l\) and computing the variance of the number of eigen-energies in these intervals. The level number variance varies linearly with $l$, $\Sigma^2(l) = l$, for spectra of an integrable system or uncorrelated spectrum. For a chaotic system, the number variance increases logarithmically due to strong correlations among energy levels. The number variance for chaotic systems is the same as that of the Gaussian Orthogonal ensemble (GOE), which is the standard benchmark for chaotic spectra, and is given by: $\Sigma^2(l) = \frac{2}{\pi^2} \left[ \ln(2\pi l) + \gamma_e + 1 - \frac{\pi^2}{8} \right]$, where \(\gamma_e \approx 0.5772\) is Euler's constant.

\begin{figure}[htbp]
	\centering
	\includegraphics[width=\linewidth]{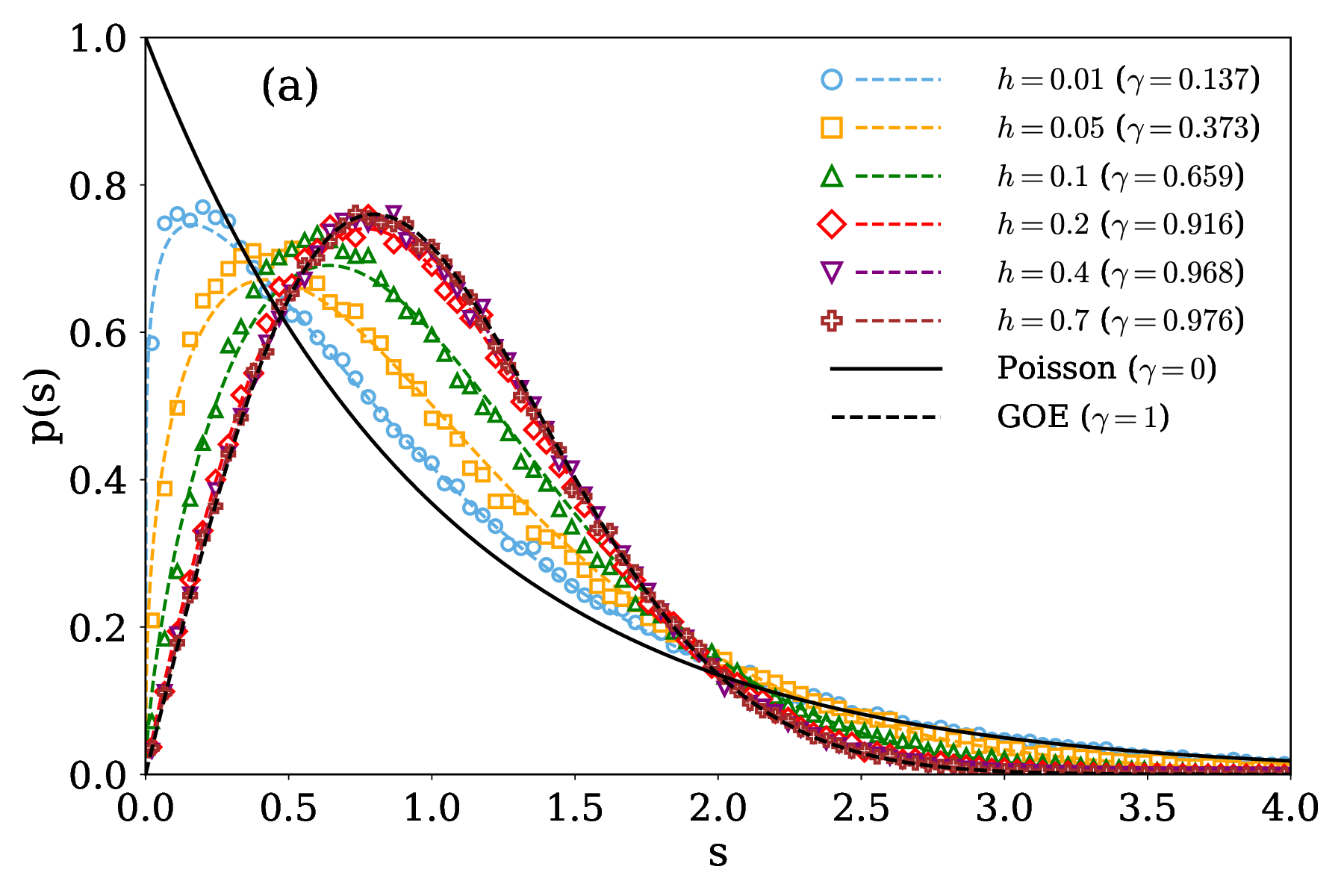}
	\includegraphics[width=\linewidth]{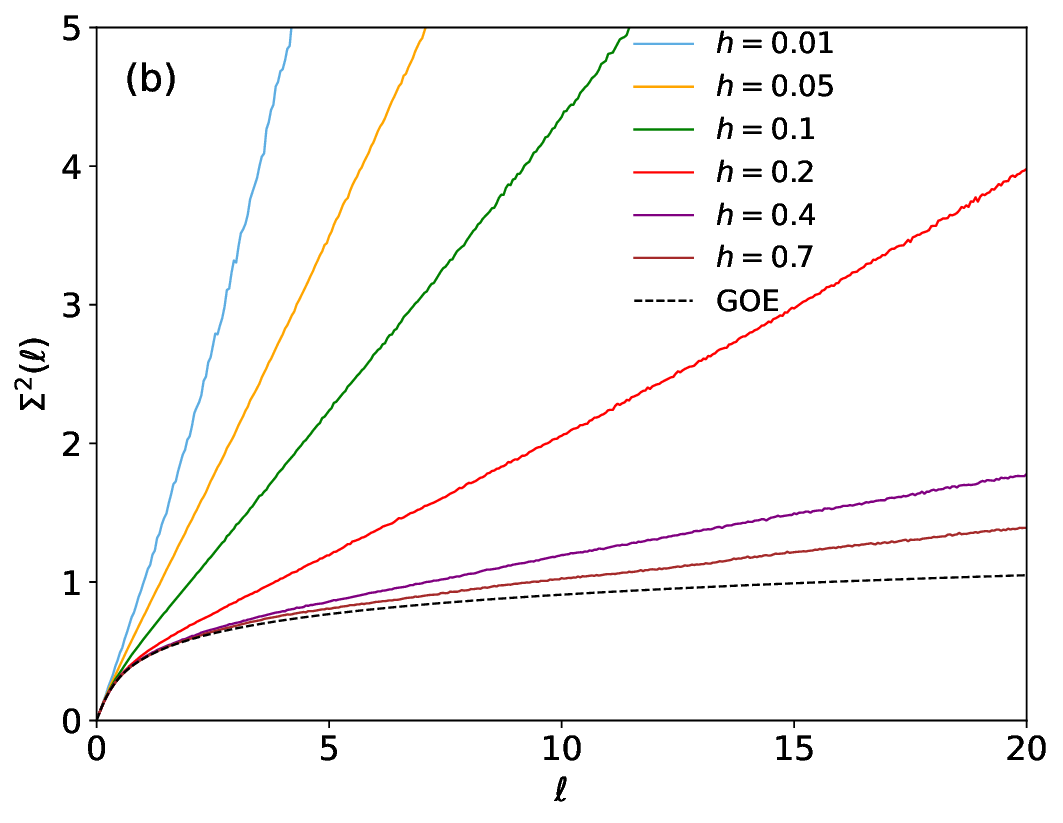}
	\caption{(Color online) (a) Nearest-neighbor spacing distribution and (b) level number variance of the XXZ Hamiltonian [see Eq.~(\ref{eq:perturbation})] across the integrability chaos crossover.}
	\label{fig:spectral_correlations}
\end{figure}

Both the fluctuation statistics for the Hamiltonian given in Eq.~(\ref{eq:perturbation}) are shown in Fig.~(\ref{fig:spectral_correlations}).
At a low field strength of $h = 0.01$, the NNSD exhibits Poisson statistics, and the number variance grows linearly, reflecting a lack of spectral rigidity. For intermediate perturbation strengths, ranging from $h = 0.05$ to $0.2$, the system enters an intermediate regime between integrable and chaotic behavior. In this region, $\Sigma^2(l)$ starts to deviate from the linear behavior and approaches the logarithmic form. At higher field strengths $h\ge0.4$, $\Sigma^2(l)$ closely aligns with the GOE prediction. A perfect alignment with the GOE prediction may not occur in many-body systems due to the locality of interactions. Note that GOE corresponds to a situation where all degrees of freedom are correlated to all other degrees of freedom.
﻿
﻿
\section{Eigenstate Thermalization Hypothesis}
\label{sec:eth}
We consider obsevables $T$ and $O$, defined as:
\begin{align}
	T &= \frac{1}{L} \sum_{i=1}^{L-1}
	\left( S_i^x S_{i+1}^x + S_i^y S_{i+1}^y \right),
	\label{eq:local_op1} \\[4pt]
	O &= \frac{1}{L} \sum_{i=1}^{L-2}
	\left( S_i^x S_{i+2}^x + S_i^y S_{i+2}^y \right).
	\label{eq:local_op2}
\end{align}

These observables represent the nearest-neighbor (NN) and next-nearest-neighbor (NNN) interactions in the $x$- and $y$-directions, respectively. Their matrix elements in the energy eigenbasis satisfy the ETH ansatz \cite{24,25}.

\subsection{Diagonal Matrix Elements}
The diagonal elements of local operators $T$ and $O$ are plotted as functions of the normalized energy \cite{25}, defined as:
\begin{align}
	\epsilon_n = \frac{E_n - E_{\text{min}}}{E_{\text{max}} - E_{\text{min}}},
	\label{eq:normalized_energy}
\end{align}
where $E_n$ is the $n$-th energy eigenvalue, and \( E_{\text{min}} \) as well as \( E_{\text{max}} \) are the lowest and highest energy eigenvalues, respectively.
This normalization ensures consistent comparison of the diagonal matrix elements across the crossover.

\begin{figure}[htbp]
	\centering
	\begin{minipage}{\linewidth}
		\centering
		\includegraphics[width=\textwidth]{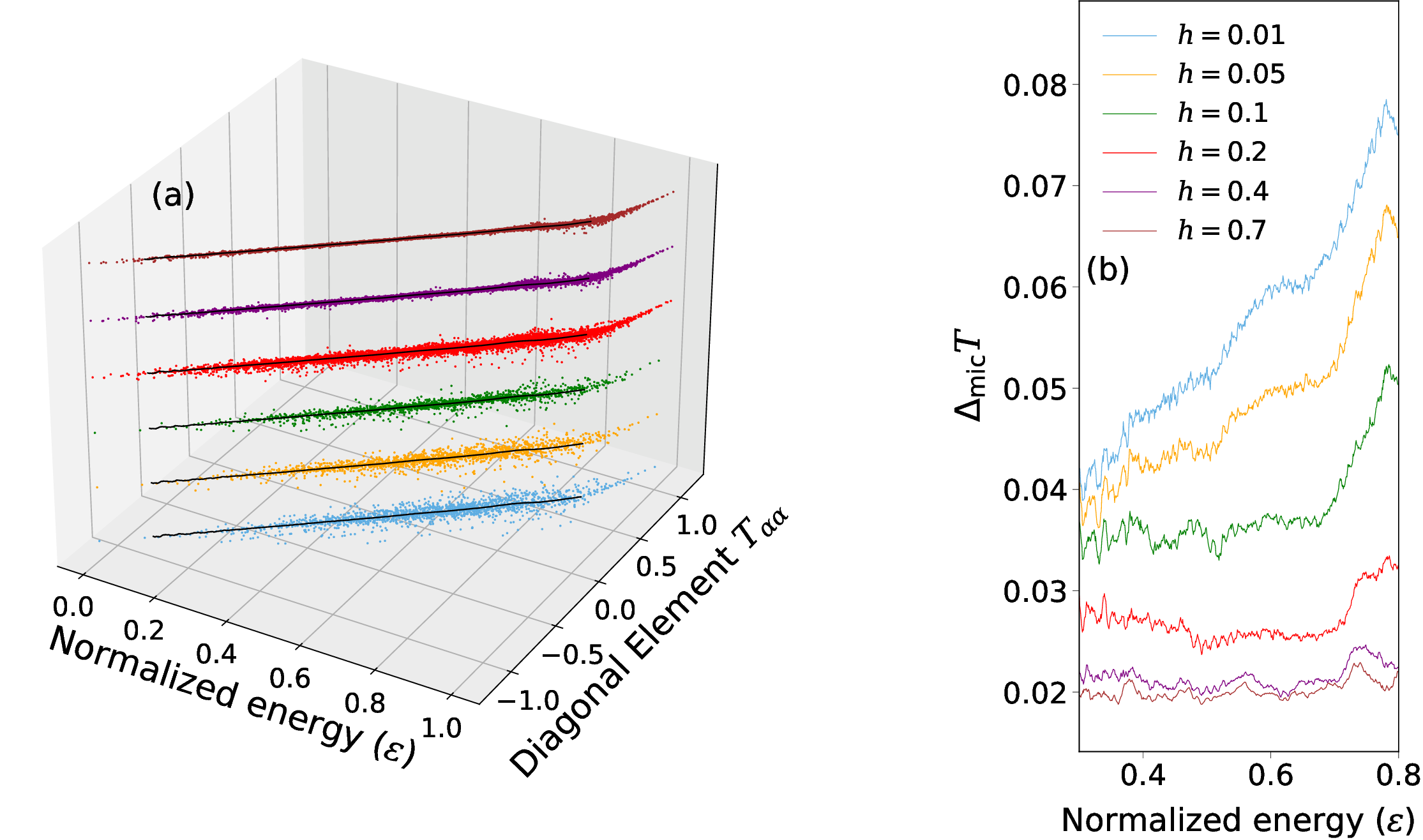}
	\end{minipage}
	\vspace{0.3cm}
	\begin{minipage}{\linewidth}
		\centering
		\includegraphics[width=\textwidth]{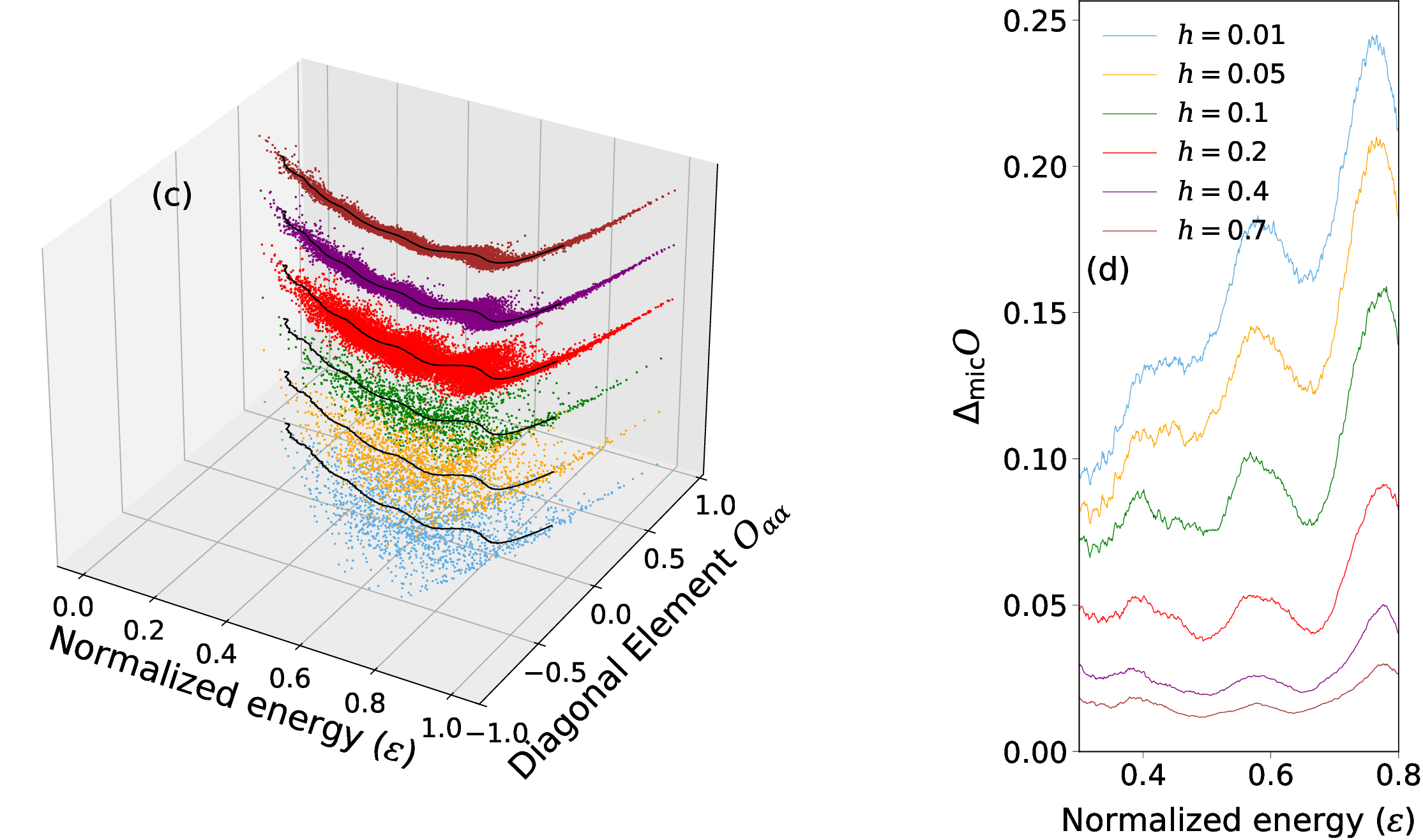}
	\end{minipage}
	\caption{(Color online) [(a),(c)] Diagonal matrix elements of the observables $T$ and $O$ with normalized energy across the crossover. The black solid line represents the microcanonical average, computed by averaging over states within \( [\epsilon_n - \delta \epsilon_n, \epsilon_n + \delta \epsilon_n] \) with \( \delta \epsilon_n = 0.02 \). [(b),(d)] Deviation \( \Delta_{\mathrm{mic}} Z \) versus normalized energy quantifying fluctuations from the microcanonical average, computed over the same energy window.}
	\label{fig:combined_operators}
\end{figure}

 In Fig. (\ref{fig:combined_operators}) we show the behavior of diagonal matrix elements of the observables $T$ Eq. (\ref{eq:local_op1}) and $O$ Eq. (\ref{eq:local_op2}) as a function of normalized energy across the crossover.
ETH predicts that \(Z_{\alpha\alpha}(E_\alpha)\) is a smooth function for chaotic systems. At the same time, integrable systems exhibit persistent eigenstate-to-eigenstate fluctuations.

For small field strengths, $ h\leq 0.1$, the diagonal elements exhibit significant fluctuations around the microcanonical average line, reflecting the system's proximity to integrability.  
As the perturbation strength increases, $ h\geq 0.4$, these fluctuations are strongly suppressed, and the diagonal elements closely follow the microcanonical average line, consistent with ETH predictions for chaotic systems. This suppression of fluctuations shows the breakdown of integrability and the emergence of thermalization.

We quantify the fluctuations of the diagonal expectation values $Z_{\alpha\alpha}$ about the microcanonical average $\overline{Z}_{\mathrm{micro}}(\epsilon_n)$, defined as \cite{55,56}:

\begin{align}
	\Delta_{\mathrm{mic}} Z(\epsilon_n) =
	\frac{1}{N(\epsilon_n)}
	\sum_{|\epsilon_\alpha - \epsilon_n|\le \delta\epsilon_n}
	\Big|Z_{\alpha\alpha} - \overline{Z}_{\mathrm{micro}}(\epsilon_n)\Big|,
\end{align}
where \(\overline{Z}_{\mathrm{micro}}(\epsilon_n)\) is the average of \(Z_{\alpha\alpha}\) over the same window of half-width \(\delta\epsilon_n = 0.02\), and \(N(\epsilon_n)\) is the number of eigenstates in this window.
This measure quantifies the extent to which individual eigenstates reproduce thermal expectations and therefore serves as a direct diagnostic of ETH behavior across the crossover.
﻿

\subsection{Distribution of Off-Diagonal Matrix Elements}

\begin{figure}[htbp]
	\centering
	\includegraphics[width=\linewidth]{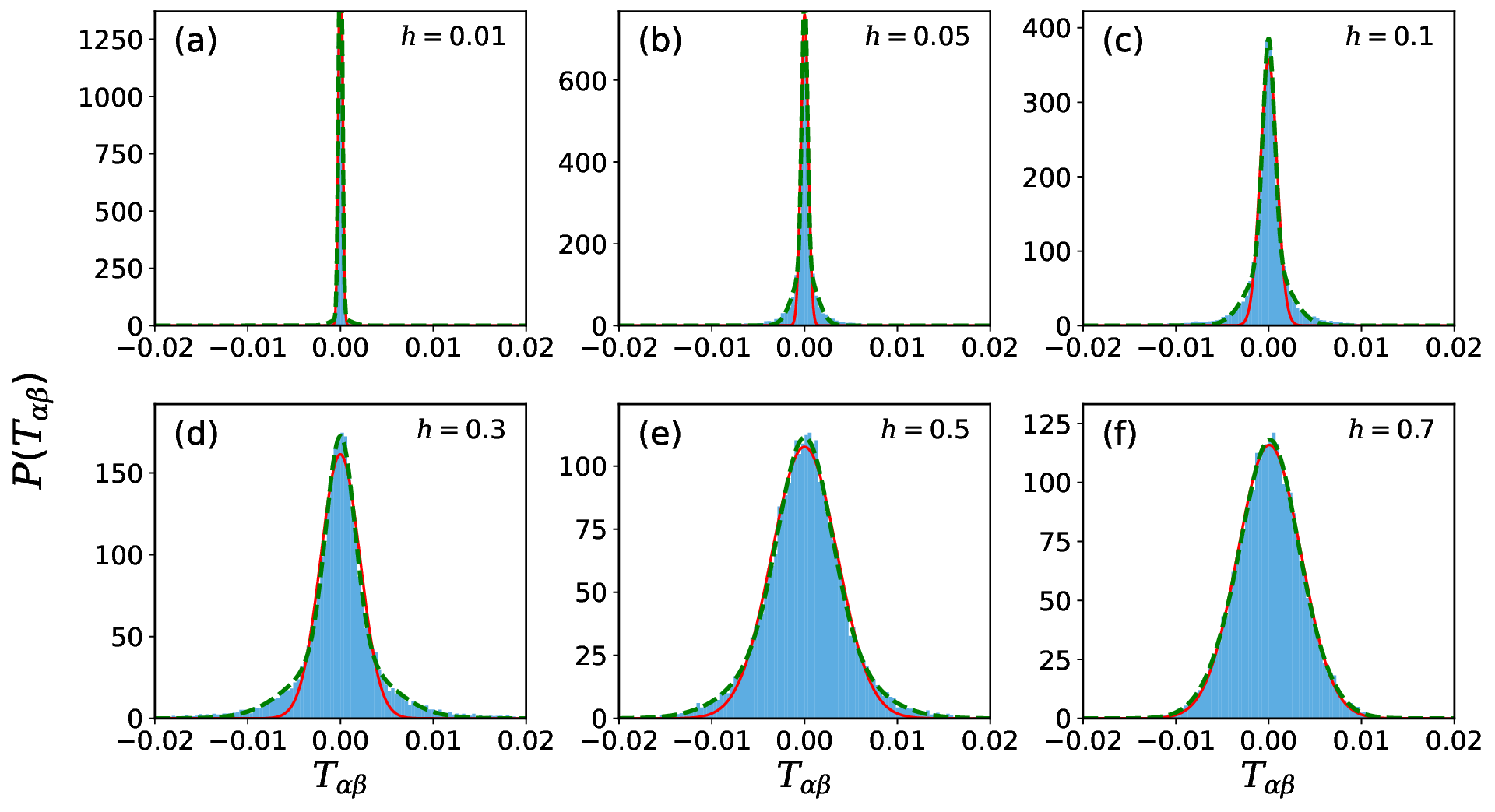}
	\includegraphics[width=\linewidth]{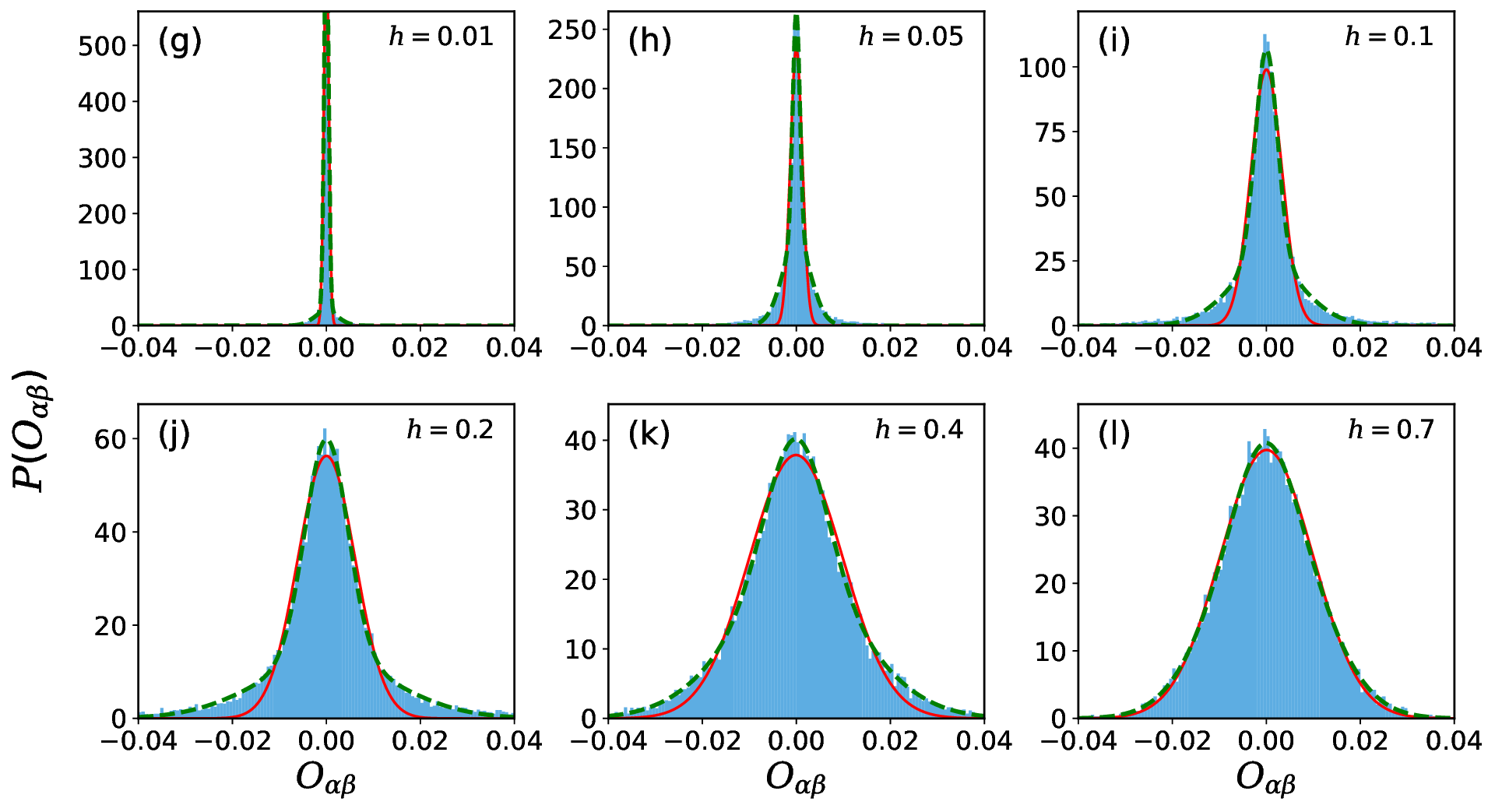}
	\caption{(Color online) Distribution of off-diagonal matrix elements \( T_{\alpha\beta} \) [panels (a)--(f)] and \( O_{\alpha\beta} \) (\( \alpha \neq \beta \)) [panels (g)--(l)] for different field strengths. The dashed green line represents a double-Gaussian fit, while the solid red line corresponds to a single-Gaussian fit.}
	\label{fig:off_diagonal_distributions}
\end{figure}

We consider the off-diagonal matrix elements of the observables $T_{\alpha\beta}$ Eq. (\ref{eq:local_op1}) and $O_{\alpha\beta}$, \(_{\alpha \ne \beta}\) Eq. (\ref{eq:local_op2}) for 200 pairs of eigenstates in the middle of the spectrum. The distributions shown in Fig. (\ref{fig:off_diagonal_distributions}) provide valuable insight into the crossover from integrability to chaos.

At low field strengths, the distribution exhibits a sharper peak than a simple Gaussian. We observed that the distribution can be better described as a mixture of two Gaussian functions with different variances \cite{28}. This behavior, observed prominently in near-integrable systems, notably this sharp peak and the deviation from a single Gaussian, indicates the presence of conserved quantities or approximate symmetries that persist even when the system is slightly perturbed.

We observe that a two-component Gaussian function provides a considerably better fit than a single Gaussian,  for small perturbation $h \leq 0.2$, supporting the idea that eigenstates remain grouped into distinct subspaces or symmetry sectors. These subspaces arise due to conserved or approximately conserved quantities in near-integrable systems. When a local operator $Z$ acts on an eigenstate \(|\alpha\rangle\), the resulting state \(Z|\alpha\rangle\) has strong overlap with eigenstates within the same symmetry sector and much weaker overlap with those in different symmetry sectors. This leads to a suppression of the off-diagonal elements \(Z_{\alpha\beta}\), when \(\alpha\) and \(\beta\) belong to different sectors, producing the sharp peak in the distribution [see Figs. \ref{fig:off_diagonal_distributions}(a)--(c) and (g)--(i)].

As the field strength increases $h \geq 0.4$, the two Gaussian components merge into a single Gaussian distribution  [see Figs. \ref{fig:off_diagonal_distributions}(d)--(f) and (j)--(l)]. In the fully chaotic regime, conserved quantities lose their relevance, and eigenstates no longer exhibit the subspace structure characteristic of near-integrable systems. Consequently, the off-diagonal elements of $Z$ behave like GOE matrix entries, resulting in a single Gaussian distribution.

The two-component Gaussian function is described as,
\begin{align}
	P(Z_{\alpha\beta}) &= w_1 P_1(0, \sigma_1^2) + w_2 P_2(0, \sigma_2^2),
\end{align}
where \( P_1(0, \sigma_1^2) \) and \( P_2(0, \sigma_2^2) \) are Gaussian distributions with mean 0 and \( w_1 \) and \( w_2 \) are the corresponding weights.
with \( w_1 + w_2 = 1 \) and \( \sigma_1 \ll \sigma_2 \). The narrow component (\( \sigma_1 \)) reflects suppressed matrix elements between eigenstates in distinct symmetry sectors, while the broader component (\( \sigma_2 \)) corresponds to intra-sector fluctuations. 
In such systems, local observables predominantly couple eigenstates within the same subspace, leading to large intra-subspace matrix elements (\( \sigma_2 \)) and suppressed inter-subspace elements (\( \sigma_1 \)). Therefore, the behavior of the system near the integrable regime is well captured by the two-component Gaussian function, arising from weakly mixed symmetry sectors.

\subsection{Variance Decay of Off-Diagonal Matrix Elements at Large \texorpdfstring{\(\omega\)}{omega}}
As \( T \) and \( O \) are averaged local operators, the ETH ansatz for the off-diagonal matrix elements must be modified, as outlined in~\cite{25,29}
\begin{align}
	Z_{\alpha \beta} = \frac{e^{-S(\bar{E})/2} f(\bar{E}, \omega) R_{\alpha \beta}}{\sqrt{L}}.
\end{align}
In quantum chaotic systems governed by the ETH, the off-diagonal matrix elements of observables represented in the eigenbasis exhibit universal statistical properties. The variance of off-diagonal elements \(\mathrm{Var}(Z_{\alpha\beta})\) serves as a key diagnostic of chaos. It quantifies the fluctuation between eigenstates \(|\alpha\rangle\) and \(|\beta\rangle\) separated by energy \(\omega = E_{\alpha} - E_{\beta}\)~\cite{23,25,28,29,54}.
Assuming \(\overline{Z_{\alpha\beta}} = 0\), which holds with high accuracy.
Now the variance of off-diagonal elements of the local operator $Z$ is expressed as:
\begin{equation}
	\overline{|Z_{\alpha\beta}|^2} \propto e^{-S(\bar{E})} |f(\bar{E}, \omega)|^2, 
\end{equation}
where \(S(\bar{E})\) is the thermodynamic entropy at mean energy \(\bar{E}\), we focus in the infinite-temperature regime, where mean energy \(\bar{E} \approx 0\) and \(S(\bar{E}) = \ln D\), and \(D\) is the Hilbert space dimension of the symmetry sector. The prefactor \(e^{-S(\bar{E})} \sim 1/D\) reflects the suppression of matrix element amplitudes in chaotic systems, scaling inversely with system size.
Thus, for chaotic systems, the spectral function is given as,
\begin{equation} 
	|f(\bar{E} \approx 0, \omega)|^2 = L D \cdot \overline{|Z_{\alpha\beta}|^2}, 
\end{equation}.

The spectral function $f(\bar{E}, \omega)$ decays exponentially at large energy differences $\omega$ as
$ \lvert f(\bar{E} \approx 0, \omega) \rvert^2 \propto e^{-\eta \omega} $,
where $\eta$ parametrizes the decay rate.

This analysis shows that observables exhibit a more generic exponential decay of their off-diagonal matrix elements, which becomes apparent at large $\omega$ \cite{23,30}.
\begin{figure}[htbp]
	\centering
	\includegraphics[width=\linewidth]{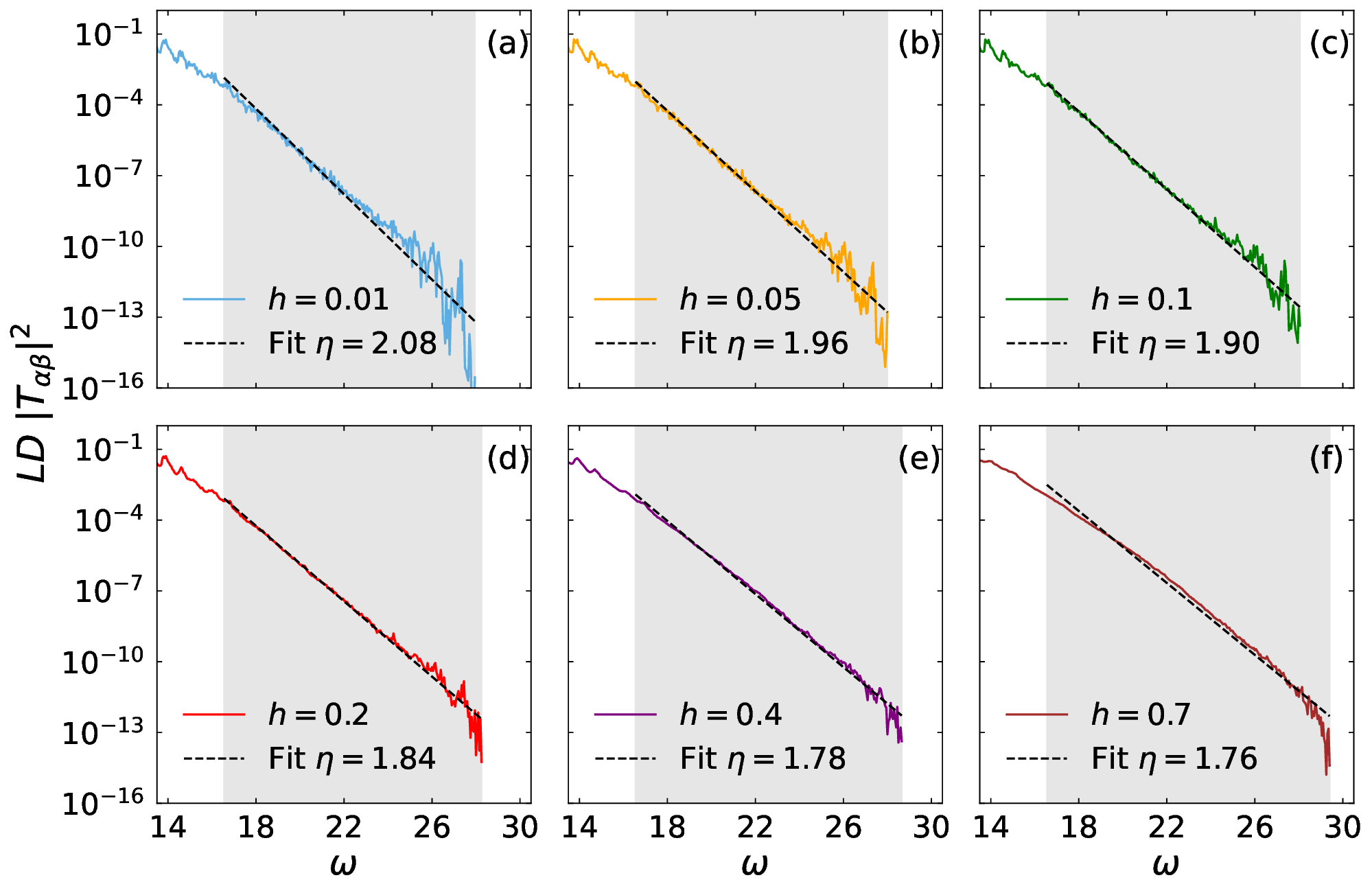}
	\includegraphics[width=\linewidth]{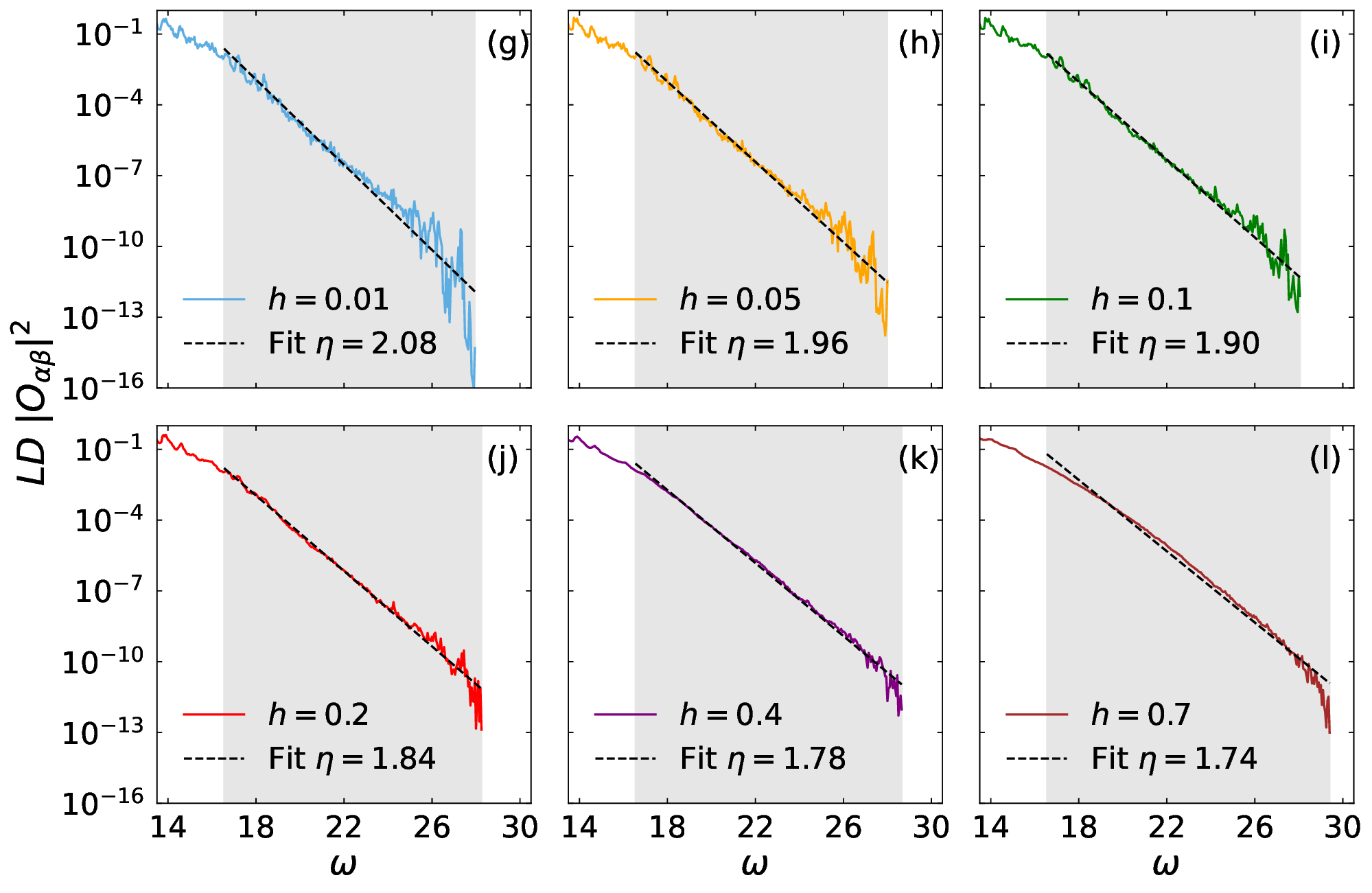}
	\caption{(Color online) Scaled variance of off-diagonal matrix elements of observables $T$  Eq. (\ref{eq:local_op1}) [(a)--(f)] and $O$ Eq. (\ref{eq:local_op2}) [(g)--(l)] versus frequency $\omega$ across the crossover. The matrix elements are computed for pairs of eigenstates whose average energy \(\bar{E} \in [-0.5, 0.5]\) and averaged over frequency bins of \(\delta \omega = 0.05\). Fitting is done in the shaded region}
	\label{fig:variance_decay_combined}
\end{figure}

The field strength $h$ controls the degree of chaos. At low $h$, minimal eigenstate mixing (weak delocalization) leads to rapid exponential decay of off-diagonal matrix elements (large \(\eta\)). Increasing \( h \) enhances mixing, allowing matrix elements to remain significant at larger energy differences \(\omega\) (smaller \(\eta\)). This manifests as a slower variance decay [see Fig. (\ref{fig:variance_decay_combined})] . This behaviour aligns with ETH, where increasing chaos promotes thermalization and smooth energy dependence of matrix elements.

Fitting the decay as \(\overline{|Z_{\alpha\beta}|^2} \propto e^{-\eta\,\omega}\), we find near-identical $\eta$ for observables $T$ and $O$. 
In the Random Matrix Theory (RMT) limit, corresponding to the fully chaotic scenario, the ETH ansatz implies that the spectral function becomes independent of both \( \omega \) and $\bar{E}$. 
Consequently, the off-diagonal matrix elements should not exhibit any frequency dependence.	Hence, in the GOE fully chaotic limit, the minimal limit of \(\eta\) is zero.
Thus, the progression from fast to slow variance decay with increasing $h$ indicates the crossover from integrability to chaos. The exponential suppression of off-diagonal elements, governed by the decay exponent $\eta$, provides a quantitative measure of this crossover.

\section{Submatrices Analysis} 
\label{sec:submatrix}
To analyse correlations among matrix elements of a local operator expressed in the eigenbasis of the XXZ model Hamiltonian [see Eq.(\ref{eq:perturbation})] across the crossover, we introduced a submatrix-based method. 
﻿

\subsection{Illustration of the Submatrices}

The Hamiltonian and observables of physical systems are fundamentally distinct from truly random operators. For instance, the matrix elements of a Pauli spin operator, such as \(\langle \alpha | \sigma_z | \beta \rangle\), must exhibit correlations to yield eigenvalues of \(\pm 1\). Consequently, the matrix elements \(R_{\alpha \beta}\) in Eq.~(\ref{Eq:ETH}) cannot be regarded as independent or uncorrelated random variables. To study these correlations in the crossover regime, we analyze submatrices of the observables.

When local observables are expressed in the energy eigenbasis, their matrix elements are predominantly distributed near the diagonal. Motivated by this structure, we extract real-symmetric blocks of size \(M\times M\) along the diagonal [see Fig. \ref{fig:subm} (a)], each serving as an energy-resolved window of the operator,
\begin{align}
	Z^{(\alpha_0)} &= \{\, Z_{\alpha_0+i,\;\alpha_0+j} \,\}_{0 \le i,j \le M},
\end{align}
where $\alpha_{0}$ represents a particular eiegenstate. Varying $\alpha_0$ allows us to probe different energy windows and corresponding diagonal blocks of $Z$.

We observe that these blocks reproduce the spectral correlations of the Hamiltonian. 

One can understand the spectral correlations of the submatrices through the lens of PLBRM. When a local observable is expressed in the energy eigenbasis, its significant matrix elements form an effective band whose width reflects the degree of the eigenstate delocalization [see Fig. \ref{fig:subm} (b)]. In the integrable regime, this band is narrow, resulting in weak mixing among eigenstates and a Poisson-like spectral statistics in the submatrices. As the system approaches the chaotic regime, the effective bandwidth increases, allowing matrix elements to couple more distant eigenstates, and the resulting submatrices begin to exhibit the spectral characteristics of the GOE. In this way, the Poisson to Wigner crossover exhibited by the submatrices directly mirrors the evolving structure of the eigenstates, ensuring that the spectral correlations encoded in the eigenstates of the Hailtonian are faithfully manifested even in the crossover regime.

Increasing the submatrix size includes matrix elements that are coupled to more distant eigenstates, and their amplitude depends on the system's chaoticity. We observe that higher-order spectral correlations also converge to those of the Hamiltonian as the submatrix size increases, indicating that long-range spectral correlations emerge as the submatrix size is enlarged [see Fig. \ref{fig:SFF_combined}, panels (g)--(l)].

This analysis of block submatrices enhances our understanding of how the matrix elements of local operators evolve across the integrability to chaos crossover.

﻿
\subsection{Statistical Properties of Submatrices: Ratio of Variances Diagonal to Off-Diagonal}
\label{sec:Statistical Properties of Submatrices: Ratio of Variances Diagonal to Off-Diagonal}

We next examine a statistical hallmark of the GOE. According to the ETH, the matrix elements of local operators in the Hamiltonian eigenbasis should exhibit GOE-like statistics. To test this, we compute the ratio of variances between diagonal and off-diagonal elements of submatrices of the observables across the crossover.

The ratio of the variances of the diagonal and off-diagonal elements of the operator $Z$ is defined as
 $R = \frac{\operatorname{Var}(Z_{\alpha\alpha})}{\operatorname{Var}(Z_{\alpha\beta})}$,
where \(\mathrm{Var}(Z_{\alpha\alpha})\) and \(\mathrm{Var}(Z_{\alpha\beta})\) denote the variances of the diagonal (\(Z_{\alpha\alpha}\)) and off-diagonal (\(Z_{\alpha\beta}\), \(\alpha \neq \beta\)) elements, respectively.

\begin{figure}[htbp]
	\centering
	
	\includegraphics[width=\linewidth]{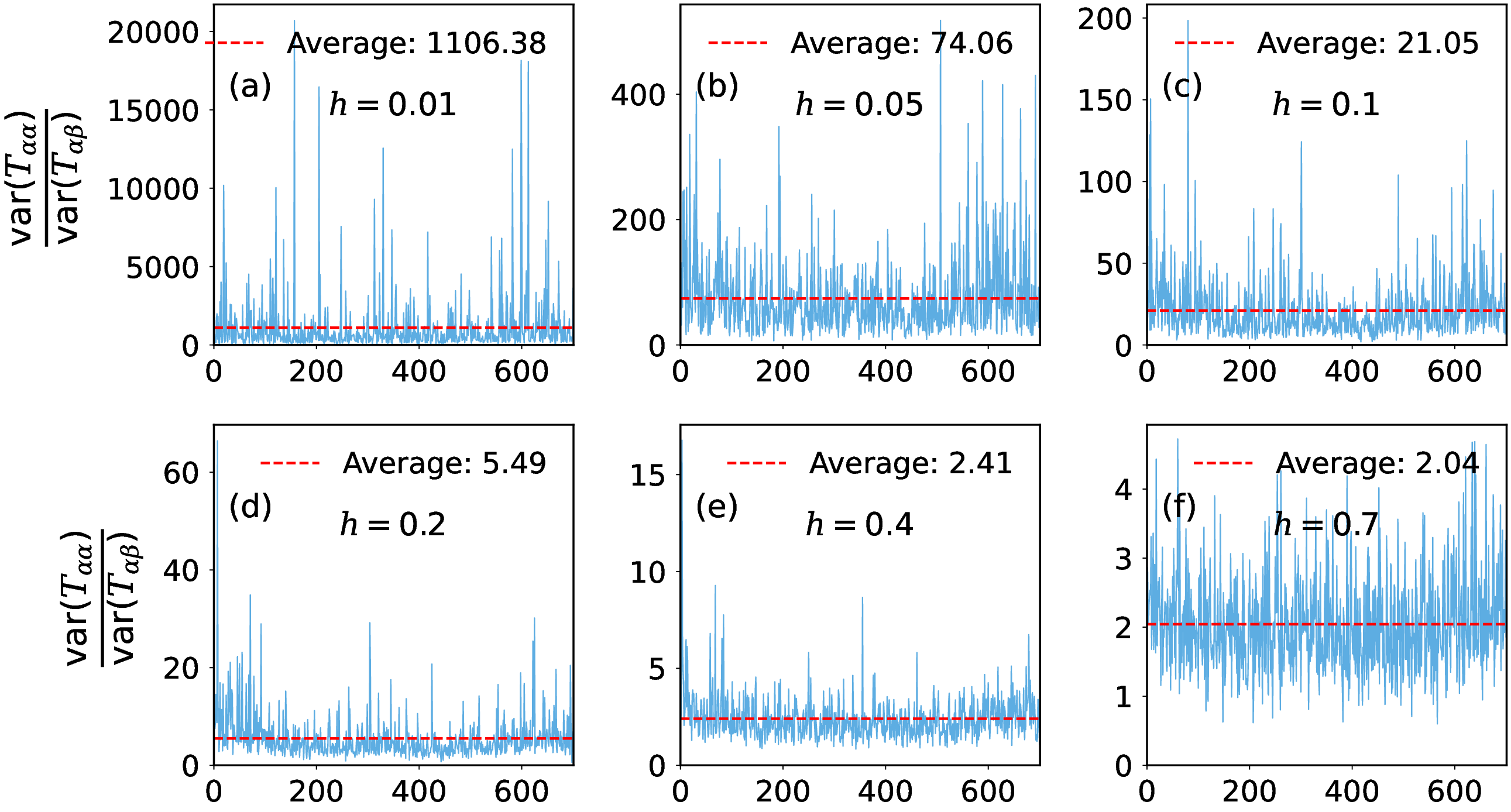}
	\vspace{0.1cm}
	\includegraphics[width=\linewidth]{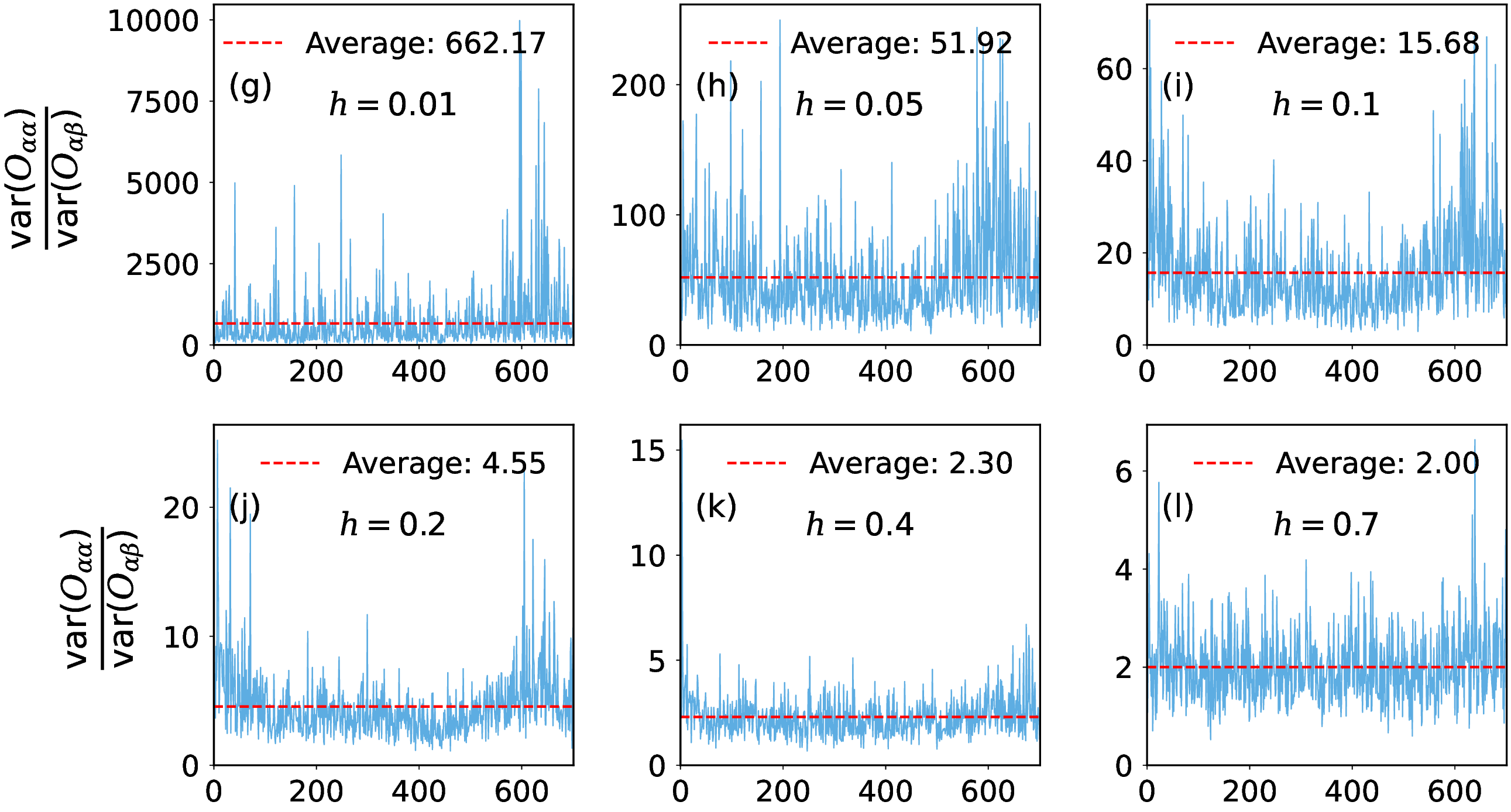}	﻿
	\caption{Variance ratio of diagonal to off-diagonal matrix elements for \(21 \times 21\) submatrices extracted from the observables $T$ Eq. (\ref{eq:local_op1}) [(a)--(f)] and $O$ Eq. (\ref{eq:local_op2}) [(g)--(l)] in the Hamiltonian's eigenbasis across the integrability chaos crossover. The $x$-axis denotes the submatrix index (1 to 700), and the $y$-axis shows the corresponding variance ratio.}
	\label{fig:variance_ratios_combined}
\end{figure}
﻿
 In Fig. (\ref{fig:variance_ratios_combined}), we show that at low \(h\), \(R\) significantly exceeds 2, reflecting the dominance of conserved quantities in the near-integrable regime. 
As $h$ increases, the system exhibits a chaotic nature, and $R$ converges to 2, aligning with GOE predictions, where the variances of diagonal and off-diagonal elements maintain a fixed ratio of 2 \cite{21,26,56}. This convergence at large $h$ highlights the breakdown of integrability and the onset of universal chaotic behavior governed by random matrix theory (RMT).

This shift illustrates how growing perturbations break down the conserved quantities defining integrable systems, promoting a statistical balance between diagonal and off-diagonal elements characteristic of chaotic systems.

\subsection {Distribution of Ratio of Consecutive Level Spacing}

The crossovers in the Hamiltonian $H$ Eq.(\ref{eq:perturbation}) by tuning the perturbation can be captured by the spectral properties of the submatrices of the observables We examine the distribution of the ratio of consecutive level spacings \cite{57,58}, which quantifies relative gaps between neighboring eigenvalues and serves as a sensitive probe of short-range spectral correlations.

To analyze the spacing-ratio distribution in local operators, we extract 700 real-symmetric submatrices of size \(21 \times 21\) along the diagonal [see Fig. \ref{fig:subm} (a)] from the local operators $T$ and $O$, expressed in the eigenbasis of the Hamiltonian $H$. By excluding boundary effects near the spectral edges, these submatrices capture correlations among matrix elements within narrow energy windows.

We find that the consecutive level spacing-ratio distributions remain unchanged with the submatrix size; however, increasing the submatrix size reduces statistical fluctuations and leads to closer agreement with the full Hamiltonian results. In contrast, the variance ratio of diagonal to off-diagonal matrix elements changes significantly. We compute the eigenvalues \(E_i\) of each submatrix and analyze the averaged distribution of spacing ratios, excluding edge eigenvalues, defined as:

\begin{align}
	r_i &= \frac{\min(s_i, s_{i+1})}{\max(s_i, s_{i+1})}, \quad \text{where} \quad s_i = E_{i+1} - E_i,
\end{align}
and $s_i$ is the gap between consecutive eigenvalues. The ratio $r_i \in [0, 1]$ eliminates the need for spectral unfolding, making it accessible for small spectrum size.

For integrable systems with uncorrelated eigenvalues, the spacing ratio distribution is characterized by Poisson statistics: $P_{\mathrm{Poi}}(r)=\frac{2}{(1+r)^2}$, For chaotic systems described by the Gaussian Orthogonal Ensemble (GOE), $P_{\mathrm{GOE}}(r)=\frac{27}{4}\,\frac{2(r+r^2)}{(1+r+r^2)^{5/2}}$,

\begin{figure*}[htbp]
	\centering
	
	\includegraphics[width=\linewidth]{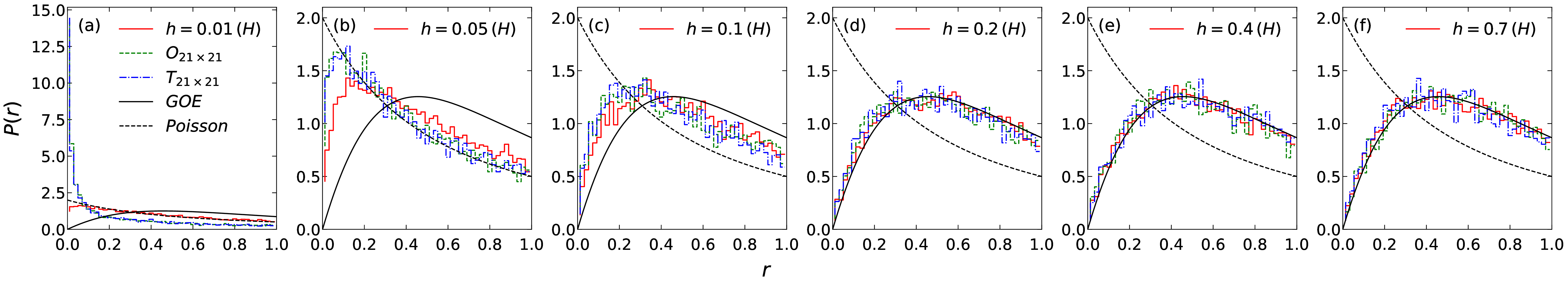}
	\includegraphics[width=\linewidth]{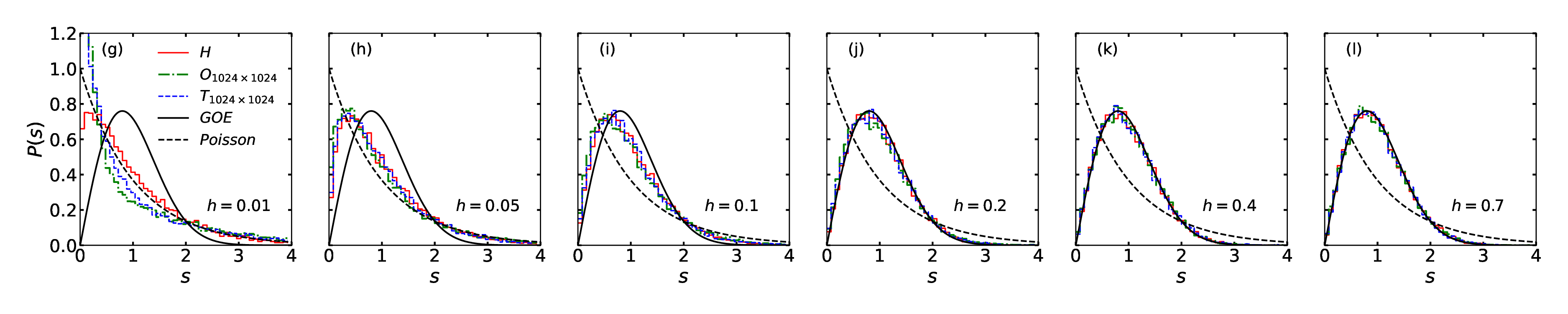}
	\caption{(Color online) Top panel [(a)--(f)] the distribution of consecutive level spacing for the XXZ Hamiltonian [see Eq. (\ref{eq:perturbation})] and for \(21 \times 21\) submatrices (700 in total) of the observables $T$ Eq. (\ref{eq:local_op1}) and $O$ Eq. (\ref{eq:local_op2}), evaluated across the crossover. Although the smallest submatrix size is considered here, increasing the submatrix size leads to closer agreement with the spectral statistics of the Hamiltonian due to reduced fluctuations in the submatrix eigenvalue density. Bottom panel [(g)--(l)] NNSD for Hamiltonian and for \(1024 \times 1024\) submatrices (16 in total) of the observables $T$ and $O$. In the legend, light red denotes the Hamiltonian $H$, blue and green denote the submatrices of operators $T$ and $O$, respectively.}
	\label{fig:submatrix_spacing_HTO}
	\vspace{0.3\baselineskip}
\end{figure*}

The distribution of the consecutive level spacing ratios for the submatrices of operators $T$ and $O$ is shown in Figure~\ref{fig:submatrix_spacing_HTO}. At weak fields (\(h \leq 0.05\)), the distribution matches \(P_{\text{Poi}}(r)\), indicating a near-integrable system with minimal mixing of eigenstates, due to the presence of the conserved quantities. As \(h\) increases, the distribution shifts toward \(P_{\text{GOE}}(r)\), becoming prominent by \(h \geq 0.4\), signaling the onset of quantum chaos with enhanced eigenstate mixing.
The GOE-like spacing ratios at high \(h\) fully support ETH's predictions for chaotic systems.
Operator $T$ and $O$ [see Eq.(\ref{eq:local_op1}) and (\ref{eq:local_op2})], exhibit similar qualitative behavior.

This crossover from Poissonian to GOE statistics reflects the breakdown of integrability and the onset of chaos. Within the ETH framework, this correspondence confirms that the matrix elements of the observables encode the same underlying spectral correlations responsible for the Hamiltonian's crossover from integrability to chaos, demonstrating that the emergence of chaos is intrinsically embedded within the operator structure itself.

To leave no doubt about the generality of this result, we have further verified the same correspondence in a distinct interacting many-body system [see Appendix. \ref{app:bh_spacing_ratio}].

\subsection{Spectral Form Factor (SFF)}

The spectral form factor (SFF) is a powerful diagnostic for probing both short and long-range spectral correlations in quantum systems \cite{36,45,46,47,48,59,60,61,62}. It is defined as:
\begin{equation}
	\text{SFF}_k(t) = \frac{1}{N} \sum_{m,n=1}^N e^{i (E_m - E_n) t},
\end{equation}
where \( N \) is the size of the system. The ensemble-averaged SFF is given by:
\begin{equation}
	\langle \text{SFF}(t) \rangle = \frac{1}{M} \sum_{k=1}^M \frac{1}{N} \sum_{m,n=1}^N e^{i (E_{m,k} - E_{n,k}) t},
\end{equation}

where $M$ is the number of realizations or samples over which the ensemble average is computed. The index $k$ refers to different realizations of the system, and the sums are taken over all eigenvalue pairs $m$ and $n$ for each realization.

In integrable systems, the SFF displays a plateau without a ramp. In contrast, chaotic systems exhibit a characteristic dip followed by a linear ramp \cite{67,68}. The dip indicates strengthening eigenvalue correlations, while the ramp signifies the onset of random-matrix-like behavior. For $h \leq 0.1$, the ramp is not clearly pronounced, indicating weaker chaotic signatures. However, as $h \geq 0.2$, both the dip and ramp become distinctly evident, reflecting robust chaotic behavior. In the long-time limit, the SFF saturates about one in both integrable and chaotic systems.

\begin{figure}[htbp]
	\centering
	\includegraphics[width=\linewidth]{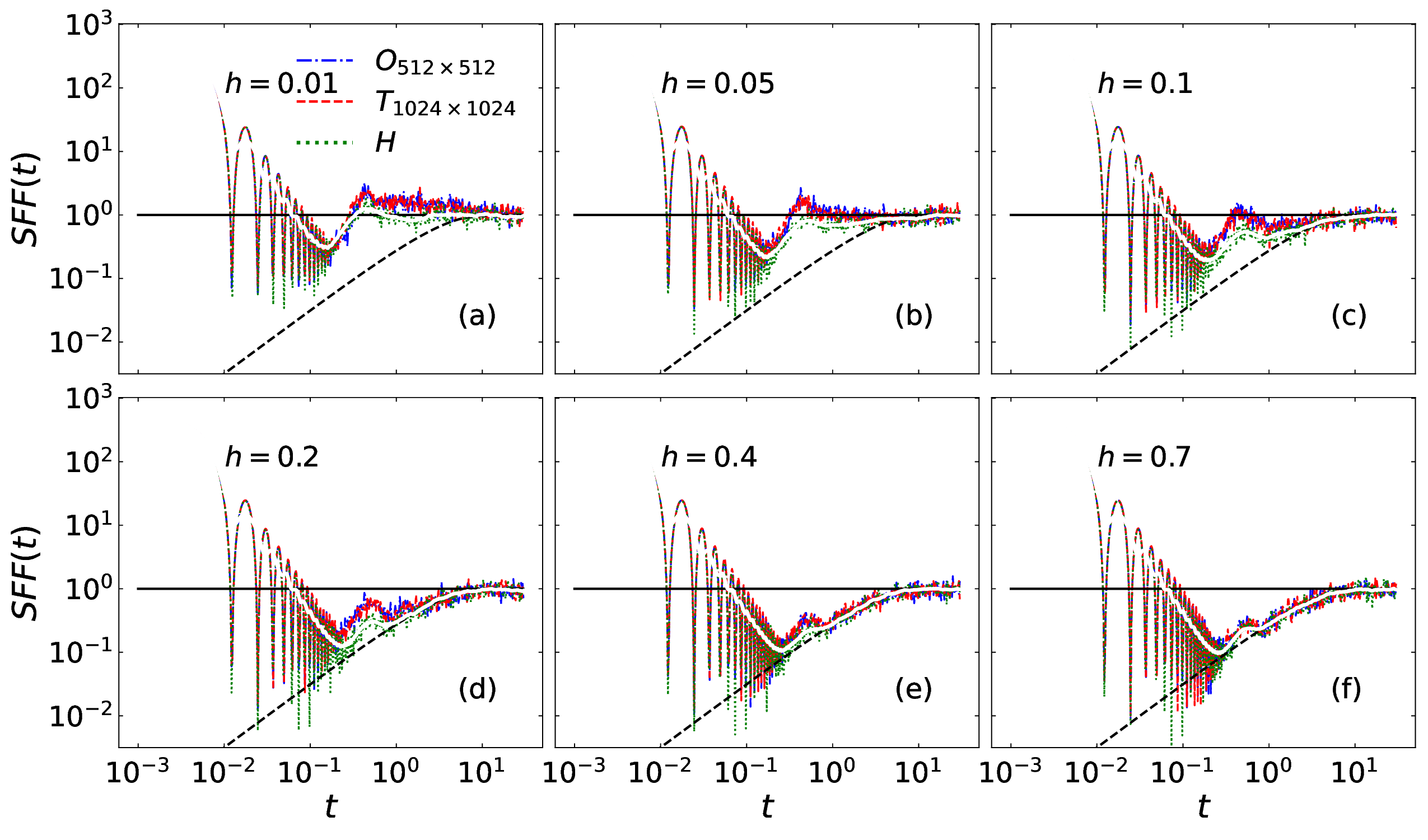}
	\includegraphics[width=\linewidth]{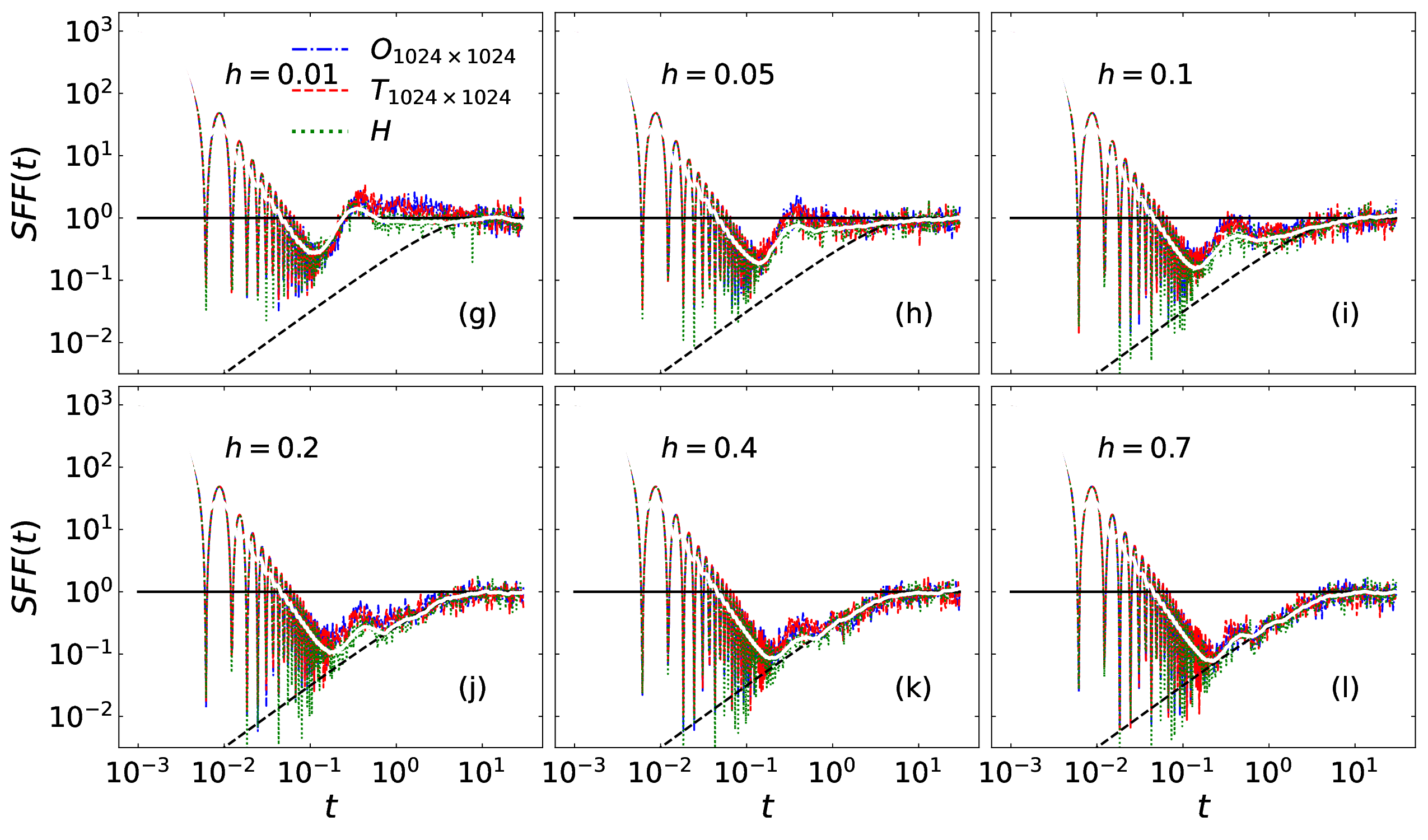}
	\caption{(Color online) Spectral form factor (SFF) for the XXZ model [see Eq.~(\ref{eq:perturbation})] and for the observables $T$ Eq. (\ref{eq:local_op1}) and $O$ Eq. (\ref{eq:local_op2}). Top panels (a)--(f) show the SFF computed over 32 submatrices of size $(512 \times 512)$ of the operators $T$ and $O$. Bottom panel (g)--(l) show the corresponding SFF obtained from 16 submatrices of size $(1024 \times 1024)$. We observe that increasing the size of the submatrices improves the overlap between the Hamiltonian and submatrix SFF data.
		Since the dimension of the full Hamiltonian is $16384$, we construct an effective ensemble by partitioning the spectrum into contiguous energy windows whose sizes match the chosen submatrix dimensions	$(512 \times 512)$ and $(1024 \times 1024)$. The spectral form factor is then averaged over these unfolded eigenvalues. As the local spectral statistics are expected to be uniform throughout the spectrum, this procedure faithfully captures the universal spectral correlations of the full system after unfolding. The black dashed line represents the RMT prediction.}
	\label{fig:SFF_combined}
	
\end{figure}

Fig. \ref{fig:SFF_combined} illustrates the SFF for the XXZ model [Eq.~(\ref{eq:perturbation})] and for submatrices of the observables $T$ and $O$ across the crossover. At low perturbation strength, no clear ramp appears in the submatrices, indicating minimal spectral rigidity. For $h \geq 0.2$ , although the short-range spectral correlations are already indistinguishable, as evidenced by the spacing-ratio statistics [see Figs. \ref{fig:submatrix_spacing_HTO}(d)--(f)]
, the SFF is not identical: the slope of the ramp increases with perturbation strength. This demonstrates that operator submatrices retain not only short-range but also long-range spectral correlations.
﻿
\subsection{Entanglement Entropy of Eigenstates}
\label{sec:entanglement_submatrix}
﻿
ETH and entanglement entropy \cite{30,63,64,65} are closely related concepts in quantum many-body systems. ETH suggests that eigenstates of quantum chaotic systems resemble thermal states, and the entanglement entropy of those eigenstates quantifies this resemblance. We have divided our system into two subsystems, \(S_A\) and \(S_B\), creating a bipartite system. We calculate the average entanglement entropy \( \langle S_A \rangle \) for subsystem \( S_A \) with sizes \( L_A = 0, 1, 2, \ldots, 14 \) across the crossover. The resulting Page curves are normalized by the maximum value, \(S_{A,\text{max}} = 7 \ln 2\), allowing us to examine how they depend on subsystem size and the degree of chaoticity in the system.
﻿
The average entanglement entropy \(\langle S_{A_,\text{ran}} \rangle\) for random pure states is given by \cite{30,42}:
\begin{equation}
	\langle S_{A_,\text{ran}} \rangle = L_A \ln 2 - \frac{1}{2} \cdot 2^{2 L_A - L}.
	\label{eq.randomEE}
\end{equation}  
﻿
﻿
To examine the correspondence between eigenstates of the submatrix extracted from the observables $T$ and $O$ expressed in the energy eigenbasis and the Hamiltonian Eq. (\ref{eq:perturbation}), we calculate the entanglement entropy across field strengths $h = 0.01$ to $0.7$, by treating the submatrix to a $(1024 \times 1024)$ dimension as the Hamiltonian describing the interaction between fictitious spin-$1/2$ sysrem.

\begin{figure}[htbp]
	\centering
	\includegraphics[width=\linewidth]{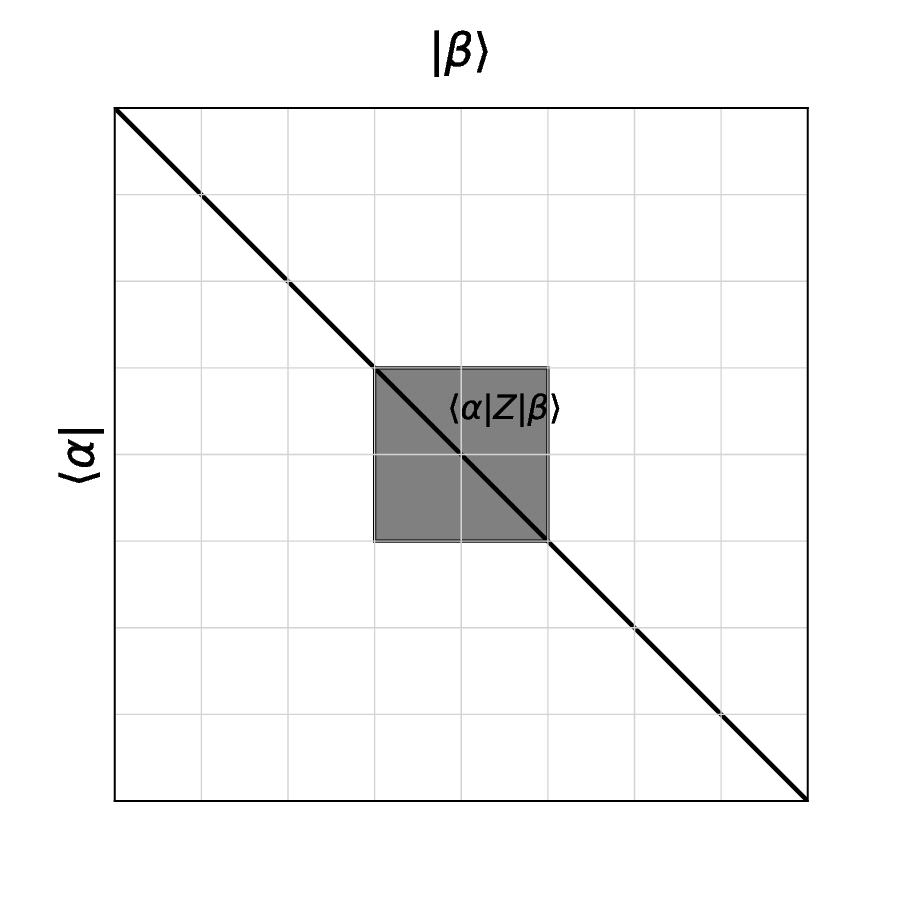}
	\caption{Schematic visualization of an operator $Z$, with the central \(1024 \times 1024\) submatrix outlined in black. The label \( \langle \alpha | Z | \beta \rangle \) denotes a representative matrix element within the submatrix.}
	\label{fig:submatrix_Z}
\end{figure}
﻿

This submatrix corresponds to mid-spectrum states, where chaotic behavior is typically most pronounced. Diagonalizing it yields \(1024\) eigenstates (effective dimension \(L_{\text{eff}} = 10\)). Each state is bipartitioned into subsystem \(A\) with size \(\tilde{L}_A = 0, 1, \ldots, 10\) and its complement \(B = 10 - \tilde{L}_A\). The resulting entanglement entropy is normalized by the maximum Page value \(S_{A,\mathrm{max}} = 5 \ln 2\).
﻿
\begin{figure}[htbp]
	\centering
	\includegraphics[width=\linewidth]{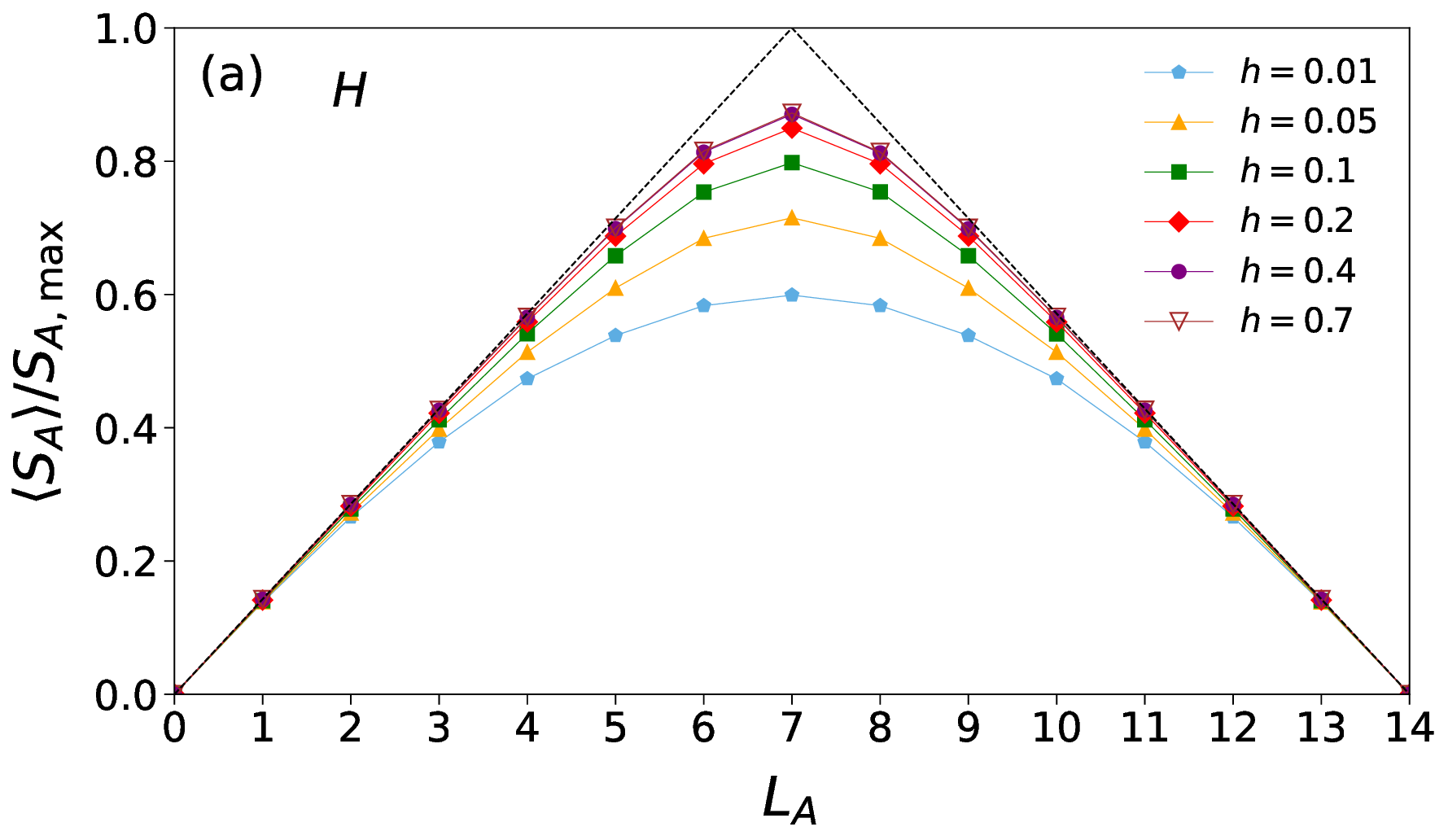}
	\includegraphics[width=\linewidth]{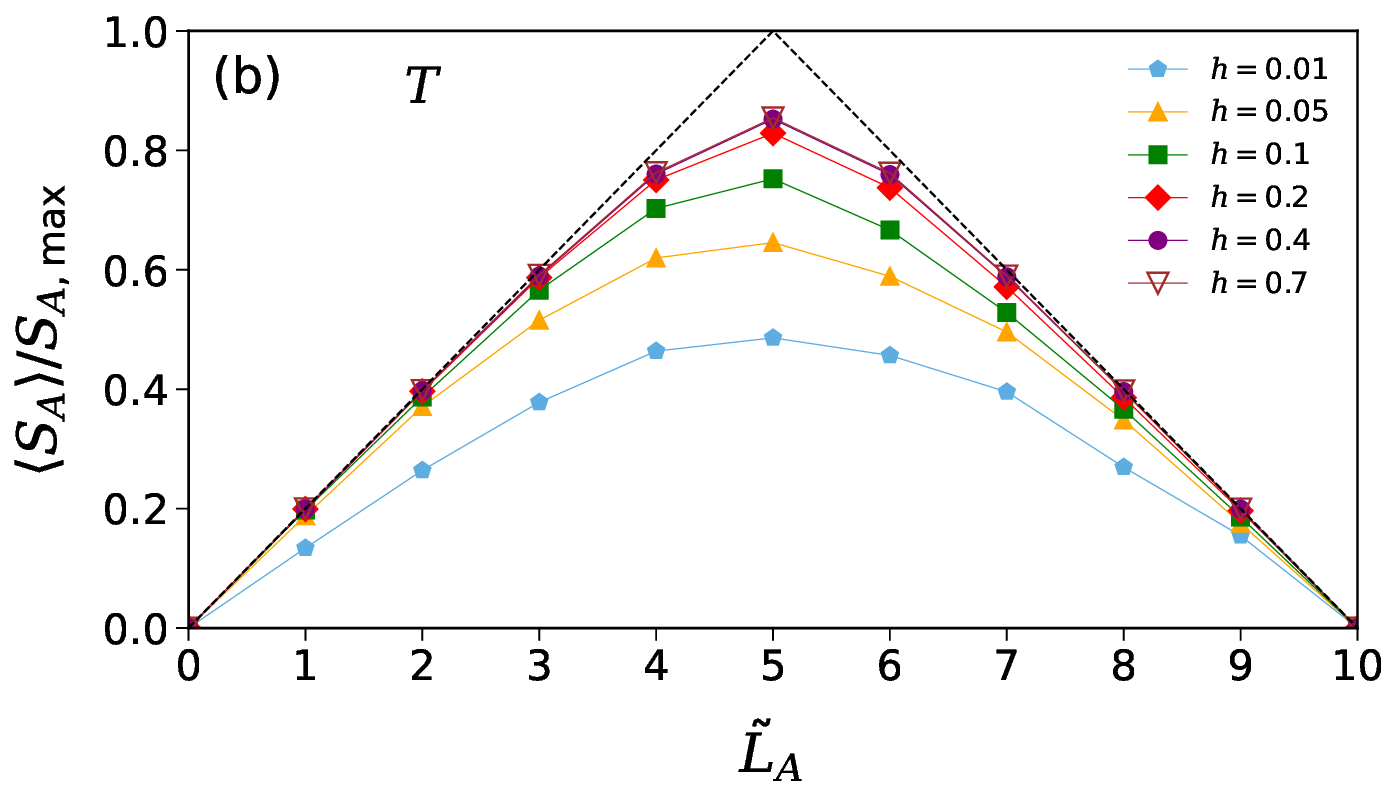}
	\includegraphics[width=\linewidth]{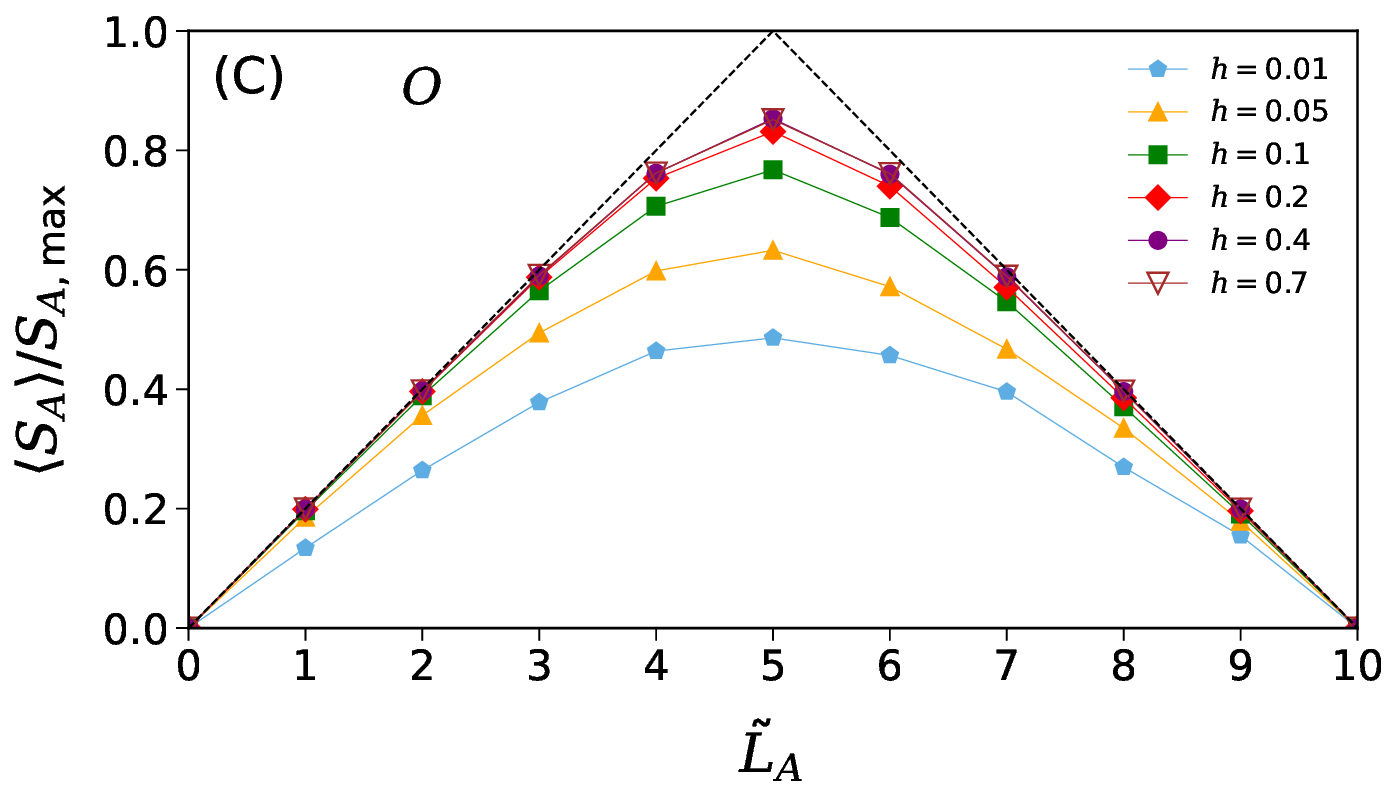}
	\caption{(Color online) Entanglement entropy $\langle S_A \rangle / S_{A,\mathrm{max}}$ as a function of subsystem size and perturbation strength.
		(a) Entanglement entropy of the XXZ model [see Eq.~(\ref{eq:perturbation})], averaged over 100 mid-spectrum eigenstates.	[(b),(c)] Entanglement entropy of a $1024 \times 1024$ submatrix of the operators $T$ and $O$, respectively, averaged over 20 mid-spectrum eigenstates.
		The dashed line denotes the random pure-state prediction Eq.~(\ref{eq.randomEE}).}

	\label{fig:entropy_submatrices_combined}
\end{figure}
﻿
Fig. \ref{fig:entropy_submatrices_combined} shows the normalized entanglement entropy of the submatrix of operators $T$ and $O$, compared with the Page curve.  
At weak fields (\(h \leq 0.1\)), the entropy lies below the Page limit, reflecting limited eigenstate mixing in the near-integrable regime.  
﻿
As the perturbation increases, the entropy grows and the curves rise toward the Page prediction, indicating enhanced mixing of submatrix eigenstates. For \(h \geq 0.2\), the entanglement profiles closely align with the Page limit, demonstrating that the submatrix eigenstates behave like thermal states. This mirrors the behavior observed in the full system, confirming that the onset of chaos and the corresponding delocalization of eigenstates are imprinted directly in the observables' structure.
﻿
Both operators $T$ and $O$ exhibit nearly the same qualitative behavior. At large (\(h \geq 0.2\)), the convergence to the Page curve is consistent with ETH expectations.
﻿
\section{summary} \label{sec:summary} 
We have investigated the crossover from integrability to chaos in a quantum many-body system, using the example of the spin-1/2 XXZ chain, by introducing a local perturbation near the middle of the site. This crossover is characterized through a comprehensive analysis of both short- and long-range spectral correlations, the ETH, and the entanglement entropy of eigenstates.

Our results show that the ETH provides a sensitive probe to analyse this crossover, as evidenced by the fluctuations of diagonal expectation around the microcanonical average [see Fig. (\ref{fig:combined_operators})] and the distributions of off-diagonal elements [see Fig. (\ref{fig:off_diagonal_distributions})] \cite{28,29}. We numerically confirm that, at large frequencies ($\omega > 16$), the variance of the off-diagonal elements of the observables decays exponentially, $\overline{|Z_{\alpha\beta}|^2} \propto e^{-\eta \omega}$, regardless of the degree of chaoticity \cite{30,54}. The decay rate systematically decreases with increasing perturbation strength [see Fig. (\ref{fig:variance_decay_combined})]. 
﻿

In our submatrix-based framework for the observables, an observable expressed in the energy eigenbasis decomposes into small real-symmetric blocks along its diagonal [see Fig. \ref{fig:subm} (a)]. These submatrices preserve the spectral statistics of the full Hamiltonian even in the intermediate regime, where the system is only partly ergodic, providing a detailed picture of how quantum chaos emerges at the operator level.

Our analysis of submatrices through level-spacing ratios [see Fig. (\ref{fig:submatrix_spacing_HTO})], spectral form factors [ see Fig. ( \ref{fig:SFF_combined})], and entanglement entropy of submatrix eigenstates [see Fig. (\ref{fig:entropy_submatrices_combined})], demonstrates clear signatures of the integrable to chaotic crossover. Remarkably, the correlations embedded in the matrix elements of the observables reproduce the same spectral correlations as the Hamiltonian. We further verified the generality of our submatrix framework in another interacting many-body system, including the many-body–localized (MBL) phase [see Appendix \ref{app:bh_spacing_ratio}], confirming its validity extends far beyond the ergodic regime. 
Furthermore, we believe that this crossover will converge even more smoothly and quantitatively as the system size increases, since a larger Hilbert space provides better statistical averaging and suppresses finite-size fluctuations.

An interesting open question is whether the correspondence revealed by our submatrix framework extends to dynamical properties, such as correlation functions and operator growth.
This work deepens our understanding of the emergence of quantum chaos in many-body systems. It provides a robust framework for future investigations into quantum many-body systems, shedding light on mechanisms driving thermalization in isolated quantum systems and paving the way for further exploration of quantum equilibration and its implications for quantum technologies.

\section*{Acknowledgments}
Shivam Mishra acknowledges the financial assistance received in the form of a Senior Research Fellowship. Numerical computations were carried out using the workstation facilities provided by the Department of Physics, Motilal Nehru National Institute of Technology (MNNIT), Allahabad, Prayagraj--211004, India.

\section*{Data Availability}
All data underlying the results presented in this work are not publicly available, but can be obtained from the authors upon reasonable request.

\appendix
﻿\setcounter{section}{0}
\renewcommand{\thesection}{A\arabic{section}}

﻿
\section{Appendix: ETH for Off-Diagonal Elements}
\label{app:matrix_elements}
We check the Gaussianity of off-diagonal matrix elements of the observables $T$ and $O$ Eq. (\ref{eq:local_op1}) and (\ref{eq:local_op2}) in the eigenbasis of the XXZ spin-1/2 Hamiltonian Eq. (\ref{eq:perturbation}) across the crossover, following Refs.~\cite{24,25,66}. 

According to the ETH, the off-diagonal elements of a local operator in the energy eigenbasis behave like entries of a GOE. To quantify this, we compute the ratio
\begin{align}
	R(\omega) = \frac{\overline{|\langle \alpha | Z | \beta \rangle|^2}}{\overline{|\langle \alpha | Z | \beta \rangle|}^2},
\end{align}

for off-diagonal elements \( Z_{\alpha\beta} = \langle \alpha | Z | \beta \rangle \) (\(\alpha \neq \beta\)) as an indicator of Gaussianity, where
\begin{align}
	\overline{|\langle \alpha | Z | \beta \rangle|^2} = \frac{1}{N_{\text{off}}} \sum_{\alpha \neq \beta} |Z_{\alpha\beta}|^2, \quad
	\overline{|\langle \alpha | Z | \beta \rangle|} = \frac{1}{N_{\text{off}}} \sum_{\alpha \neq \beta} |Z_{\alpha\beta}|,
\end{align}

and \( N_{\text{off}} \) is the number of off-diagonal elements in the energy window centered at \( \omega \).
﻿
\begin{figure}[htbp]
	\centering
	\includegraphics[width=\linewidth]{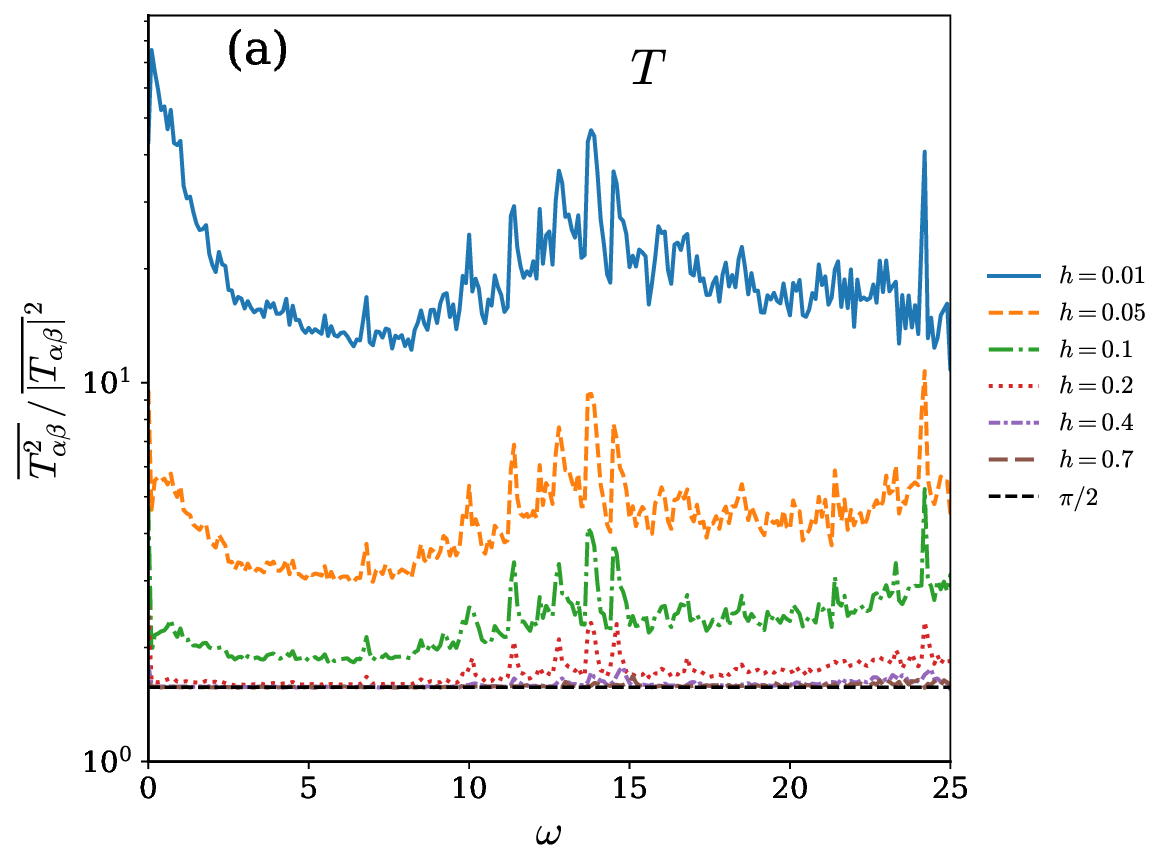}
	\includegraphics[width=\linewidth]{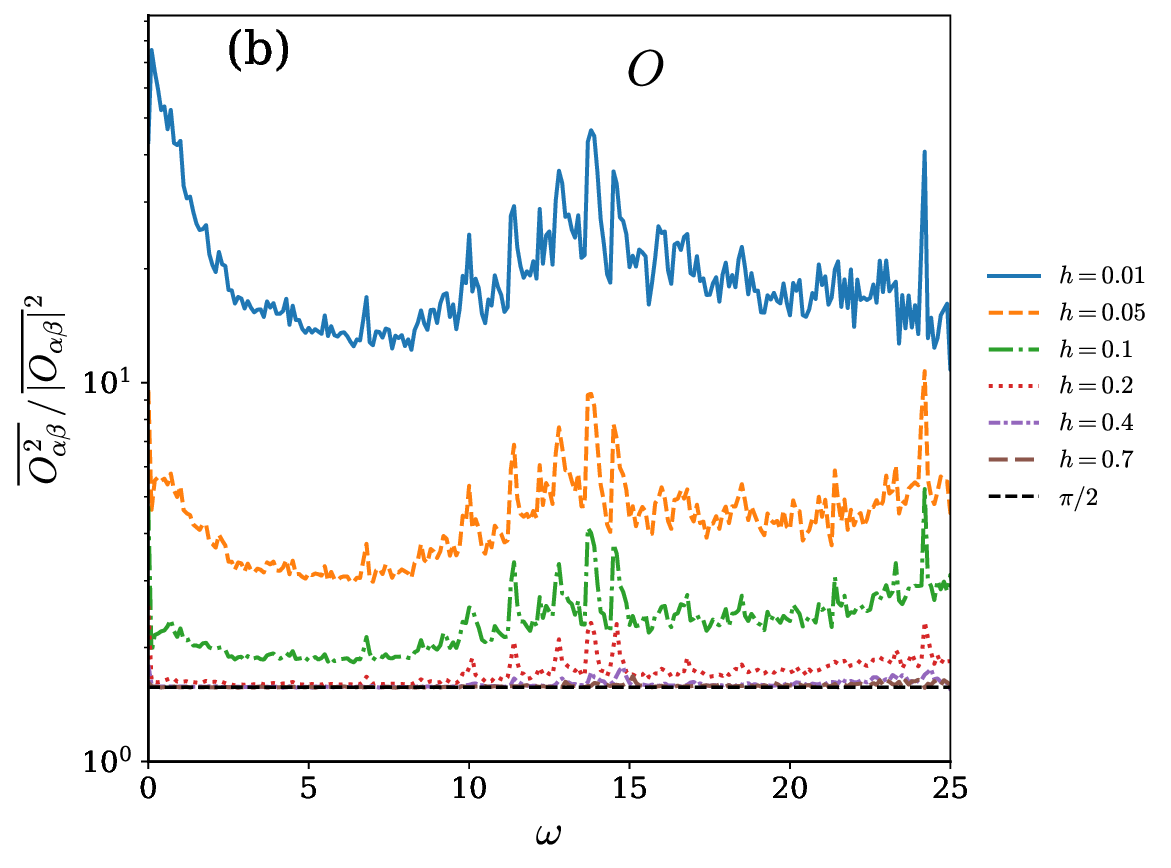}
	\caption{(Color online) Ratio \( R(\omega) \) for operators $T$ (a) and $O$ (b). The matrix elements are computed for pairs of eigenstates whose average energy \(\bar{E} \in [-0.5, 0.5]\), averaged over frequency bins of width \(\delta \omega = 0.1\). The dashed line represents the RMT prediction.}
	\label{fig:ratios_combined}
\end{figure}
﻿

Figure~\ref{fig:ratios_combined} shows the ratio \( R(\omega) \) for operators \( T \) and \( O \). 
At low perturbation strengths (\( h \leq 0.1 \)), $R(\omega)$ deviates from the RMT prediction, reflecting the presence of underlying symmetries and
minimal eigenstate mixing (weak delocalization) characteristic of the near-integrable system. As the perturbation strength increases \( h \geq 0.2 \), these symmetries are progressively broken, leading to higher delocalization of eigenstates. 
For higher values of perturbation, \( R(\omega) \) converges toward the RMT prediction $\pi/2$, indicating that the off-diagonal elements are Gaussian distributed, consistent with the  ETH ansatz.
\section{Appndix: Spacing-Ratio Distribution for Submatrices of the Bose--Hubbard Model}
\label{app:bh_spacing_ratio}
To further test the generality of our submatrix framework, we verify it for the one-dimensional disordered Bose--Hubbard model \cite{69,70} by computing the spacing-ratio distribution \( P(r) \) of eigenvalues for submatrices constructed from the local operator,
﻿
\begin{equation}
	\begin{aligned}
		H_{\mathrm{BH}} = &-J \sum_{\langle i,j\rangle} (b_i^{\dagger} b_j + \mathrm{H.c.}) 
		+ \frac{U}{2} \sum_i n_i (n_i - 1)  \\
		&+ \sum_i \epsilon_i n_i ,
	\end{aligned}
	\label{eq:bhamiltonian_mbl}
\end{equation}
where $J$ is the hopping amplitude and $b_i^\dagger$($b_j$) are the bosonic creation and annihilation operators at site $i$ and $j$,
$U$ is the on-site interaction strength. 
The weak disorder $\epsilon_i \in ( -0.05, 0.05)$ destroys all the nontrivial symmetries.

By tuning the interaction ratio $ U/J $, the spacing ratio exhibits a crossover from Poisson to Wigner, and for sufficiently strong interaction in the presence of disorder, it reverts to Poisson statistics consistent with the many-body localized (MBL) phase.

The observable used to construct the submatrices is defined as,  
\begin{align}
	O_{\mathrm{BH}} = \sum_i n_i ,
	\label{eq:obh}
\end{align}
where \( n_i = b_i^{\dagger} b_i \) represents the on-site occupation number.
﻿
Exact diagonalization was performed for a lattice of \( L = 8 \) sites with a total of \( N = 8 \) bosons; eigenstates from the edges of the spectrum were excluded from the analysis. 
Submatrices of dimension  \(50 \times 50\) were extracted from the matrix representation of the operator [see Eq.~\eqref{eq:obh}] in the energy eigenbasis, following the same construction used for the XXZ model. 

\begin{figure*}[htbp]
	\centering
	\includegraphics[width=\textwidth]{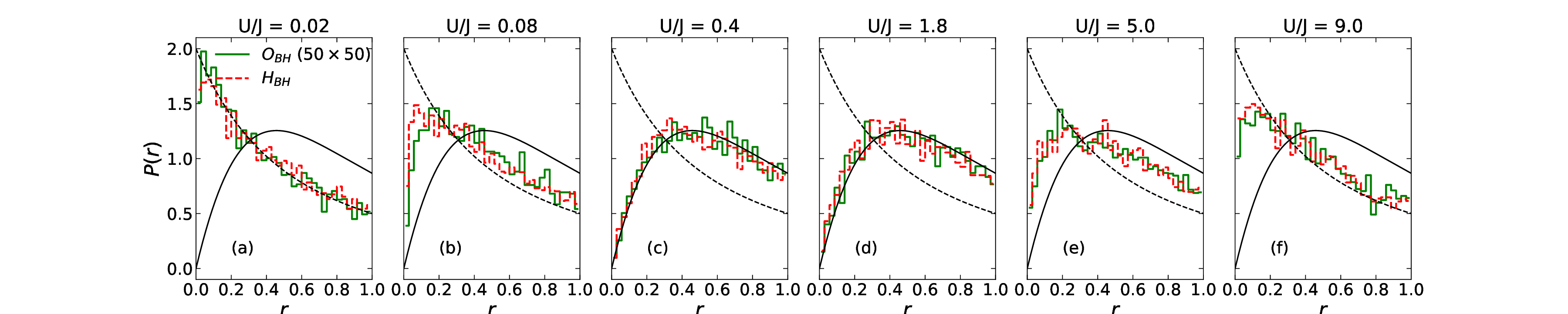}
	\caption{(Color online) The distribution of consecutive level spacing ratios for the submatrices in the one-dimensional Bose-Hubbard model. [(a)--(f)] shows the nearest-neighbor spacing ratio $P(r)$. The representative values of interaction ratio $U/J = \{0.02, 0.08, 0.4, 1.8, 5.0, 9.0\} $ cover distinct regimes of the system:  the integrable to chaotic crossover $U/J$ = (0.02, 0.08, 0.4) and the chaotic to many-body localized (MBL) crossover $U/J$ = (1.8, 5.0, 9.0). For small $ U/J$,  spacing ratio $P(r)$ follows the Poisson distribution, characteristic of integrable spectra. At an intermediate value of $U/J$, $P(r)$ approaches the GOE prediction, signaling the onset of quantum chaos. At a higher value of the interaction ratio $U/J$, the spacing ratio reverts to the Poisson distribution, consistent with the emergence of many-body localization. This evolution demonstrates that submatrix spectra faithfully reproduce the Poisson to Wigner crossover (including MBL phase) observed in the Hamiltonian.}
	\label{fig:Sp_ra_BH}
\end{figure*}

These results confirm that submatrices encode the same universal spectral correlations as the full Hamiltonian.
﻿
\section{Appendix: SFF of the Hamiltonian for Single Realization}
\label{app:single_sff}
We compute the SFF of the Hamiltonian defined in Eq. (\ref{eq:perturbation}) across the crossover regime. Notably, the SFF evaluated from a single realization already captures the same qualitative features, such as the onset of RMT behavior, as those obtained by decomposing the spectrum into several partitions [see Fig. (\ref{fig:SFF_combined})].
\begin{figure}
	\centering
	\includegraphics[width=\linewidth]{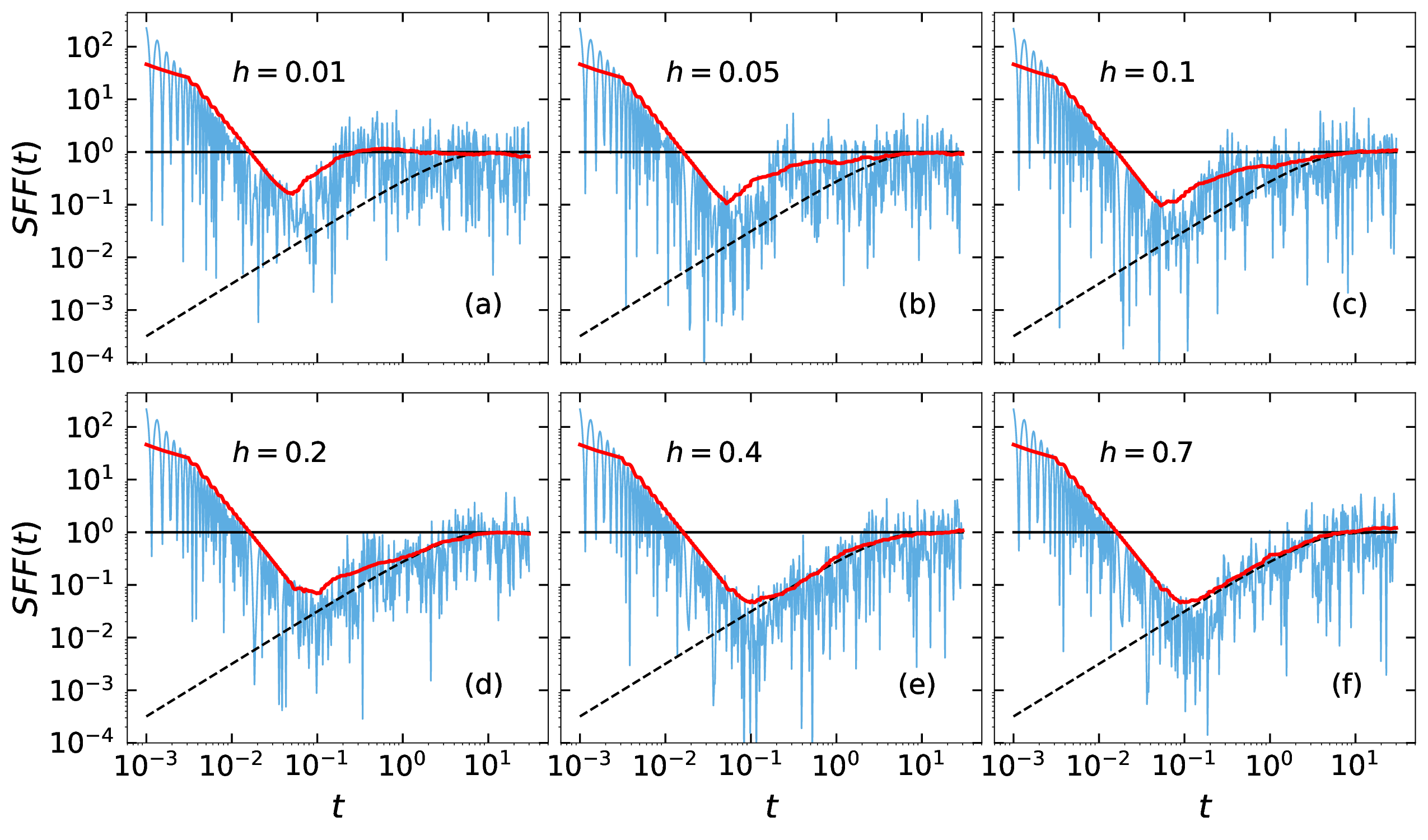}
	\caption{SFF of the Hamiltonian calculated for a single realization.
		The red curve represents the moving average used as a guide to the eye.
		Dashed and solid black lines represent the GOE and saturation predictions, respectively.}
	\label{fig:SFF_single}
\end{figure}

Figure (\ref{fig:SFF_single}) shows the single-realization spectral form factor, the overall profile is not changing significantly as compare to Fig. (\ref{fig:SFF_combined}).

\bibliography{ref}%
\end{document}